\documentclass[floatfix, reprint]{revtex4-2}

\usepackage{aps_preamble}

\begin{document}
	
	\preprint{APS/123-QED}
	
	\title{Waveform degeneracy of binary systems and Lagrange three-body systems}
	
	\author{Carlos Jaimel Doctolero}
	\email{cdoctolero@nip.upd.edu.ph}
	\affiliation{National Institute of Physics, University of the Philippines, Diliman, Quezon City 1101, Philippines}
	\author{Ian Vega}
	\email{ivega@nip.upd.edu.ph}
	\affiliation{National Institute of Physics, University of the Philippines, Diliman, Quezon City 1101, Philippines}
	
	\date{August 11, 2026}
	
	\begin{abstract}
		A particular solution to the three-body problem is the circular Lagrange three-body system, where the masses move in circular orbits such that they always constitute an equilateral triangle. Such a system has been found to emit gravitational waves with a waveform similar to that of a binary system. In this work, we study the gravitational waveform degeneracy between quasicircular binary systems and Lagrange three-body systems up to 0.5PN order. Assuming we know the parameters of a given binary system, we determine the parameters of the Lagrange triple that produces the same waveform as that of the binary. We show that there exists a mass quadrupole degeneracy in both the plus and cross modes, characterized by two parameters. We also find that there are binary systems and linearly stable Lagrange three-body systems that can have the same mass quadrupole waveform up to the coalescence time. In such cases, the normalized overlap of the waveforms with respect to the power spectral density of the advanced LIGO design remains above 0.97 as long as the binary has nearly symmetric masses. Beyond the mass quadrupole, there is a unique degeneracy at the 0.5PN. However, the Lagrange triple that satisfies this degeneracy is unstable.
	\end{abstract}
	
	\maketitle
	
	
	\section{Introduction}\label{sec:intro}
		Coalescing binary systems are promising sources of gravitational waves (GWs) as they radiate large amounts of energy via gravitational radiation. The first detection of gravitational waves in 2015 came from a binary system composed of stellar-mass black holes \cite{ligo_observation_2016}. Subsequent detections include GWs from other binary black hole systems \cite{GW190412_2020, GW190814_2020, GW170814_2017}, neutron star binary systems \cite{GW170817_2017, liu_2021}, and binary systems composed of a black hole and a compact object \cite{GW190814_2020}. These systems can form in isolation or in dense regions like globular clusters or in the centers of galaxies \cite{romero-shaw_2021, stegmann_2022}. The detection of GWs from coalescing binary systems allowed us to confirm many of the predictions of general relativity for the dynamics of binary systems.
		
		In the slow, inspiral stage of coalescing binaries, the orbital evolution and GW emission can be described using analytic perturbative methods such as the post-Newtonian (PN) approximation, which is an expansion of the equations of motion of the system in powers of $G$ and $1/c^2$ \cite{pati_2000, blanchet_2014, blanchet_2024}. The same formalism can be applied to three-body systems. However, there is an added complexity: even in Newton's theory of gravity, three-body systems are notorious for being sensitive to initial conditions and for lacking general closed-form solutions \cite{musielak_2014}. Nonetheless, there are cases of the three-body problem whose solutions are stable and have periodic orbits, making them candidates for gravitational wave studies \cite{li_2021}. Examples include Moore's stable figure eight orbit, where three bodies of equal masses all move periodically in a single figure eight \cite{moore_1993}, and Henon's crisscross \cite{henon_1976, moore_2006}. Relativistic corrections to figure eight orbits have been studied by previous authors \cite{imai_2007, li_2021}. The mass quadrupole GWs of such systems have also been studied \cite{torigoe_2009, galaviz_2011, dmitrasinovic_2014}.
		
		There are also three-body systems that can be described analytically, like the Lagrange three-body system, where three bodies of finite masses move in elliptic orbits around their center of mass (COM) so that they always constitute an equilateral triangle \cite{essen_2000, torigoe_2009, musielak_2014}. Certain combinations of masses make the system stable under small perturbations in Newtonian gravity \cite{mansilla_2006, roberts_2002, sicardy_2010}. An example of a Lagrange triple is the system formed by the Sun, Jupiter, and the Trojan asteroids \cite{gurfill_2016}. One mechanism explaining the formation of astrophysical systems in a Lagrange triangular configuration is the capture of a third body at the stable $L_4$ or $L_5$ Lagrange points of a binary; this can be assisted by resonant trapping. While a binary is still separated by large distances, gas or dust can also accumulate at the stable Lagrange points, potentially forming a star at these regions \cite{morbidelli_2005, schnittman_2010, seto_2010}. Numerical simulations also suggest that triple systems with periodic orbits can form through interactions among two binary systems or between a triple system and a fourth body \cite{zwart_2026}.
		
		Torigoe \textit{et al.} \cite{torigoe_2009} observed that the mass quadrupole GWs and frequency sweep of binaries and Lagrange triples have the same form, thereby making it difficult to distinguish the two systems using only the mass quadrupole waveforms. In a following paper, Asada \cite{asada_2009} studied how one can determine the parameters of the Lagrange triple assuming that the waveform comes from a Lagrange triple. In their work, they also established that the presence of certain observable time lags between the mass quadrupole waveform and the combined waveform of the mass octupole and current quadrupole should be an indication that the source is indeed a Lagrange three-body system. Otherwise, the gravitational waveform may be produced by other systems.
		
		There have been efforts to describe the dynamics and stability of three-body triangular configurations at higher PN orders. Ichita \textit{et al.} \cite{ichita_2011} found that an equilateral triangle configuration satisfies the 1PN equations of motion if and only if all the masses are equal. Subsequent papers showed that a triangular configuration for arbitrary masses can exist at 1PN, but it is not always equilateral due to relativistic corrections to each side, and is less stable than its Newtonian counterpart \cite{yamada_2012, yamada_2015}. Yamada \textit{et al.} \cite{yamada_2016} also showed that the gravitational radiation reaction force, which appears at 2.5PN, can adiabatically shrink the Lagrange three-body system while keeping its equilateral triangle configuration. More recently, it was shown that three masses moving in elliptic orbits can also form a triangle configuration in the parametrized post-Newtonian (PPN) formalism \cite{nakamura_2023, nakamura_2024}.
		
		Waveform similarity is crucial for parameter estimation since we need to match the strain data with a theoretical model of the waveform. Degeneracies between waveforms will lead to incorrect descriptions of the GW source. In this paper, we revisit Asada's work and provide a detailed assessment of the degeneracy they first observed. More specifically, we show analytically that there is still a waveform degeneracy between binary systems and Lagrange three-body systems, even when the time lag between the mass quadrupole waveform and the 0.5PN waveform vanishes. Given the parameters of a binary system, we show that the GWs of a binary and a stable Lagrange triple can be degenerate up to the mass quadrupole. In such cases, the system with the larger chirp mass coalesces first, and it is even possible that a stable Lagrange triple and a binary can have the same coalescence time. Beyond the mass quadrupole, there is a unique waveform degeneracy up to the 0.5PN. However, the Lagrange triple that satisfies this degeneracy is unstable.
		
		This paper is organized as follows: in Sec. \ref{sec:reviewGW}, we briefly review the linearized theory of GWs; in Sec. \ref{sec:waveforms}, we review the GWs of quasicircular binary systems and Lagrange three-body systems and provide insights about some subtleties in the waveform of the Lagrange triple; in Sec. \ref{sec:degeneracy}, we study the waveform degeneracy of the two systems; finally, in Sec. \ref{sec:conclusion}, we summarize our findings. The equations are in the units $G = c = 1$, but the plots will be in SI units.
	
	\section{Review on the linearized theory of gravitational waves}\label{sec:reviewGW}
		Let $\vu{n}$ be the unit vector pointing towards the detector. We define
		\begin{align}
			\Pi_{ij} &= \delta_{ij} - n_i n_j, \\
			\Lambda_{ijk\ell} &= \Pi_{ik} \Pi_{j\ell} - \frac{1}{2} \Pi_{ij} \Pi_{k\ell}.
		\end{align}
		The tensor $\Pi_{ij}$ is a projection tensor onto the plane orthogonal to $\vu{n}$ while $\Lambda_{ijk\ell}$ projects a tensor onto the transverse-traceless (TT) gauge. In the linearized theory of gravity, we separate the spacetime metric $g_{ab}$ as $g_{ab} = \eta_{ab} + h_{ab}$. The multipole expansion of the solution to the linearized Einstein equation is
		\begin{equation}
			h^{\mathrm{TT}}_{ij} = h^{\mathrm{TT}}_{ij,\mathrm{quad}} + h^{\mathrm{TT}}_{ij,\mathrm{oct+cq}} + \mathcal{O}(1/c^2),
		\end{equation}
		where the mass quadrupole $h^{\mathrm{TT}}_{ij,\mathrm{quad}}$ is the leading order term, and the sum $h^{\mathrm{TT}}_{ij,\mathrm{oct+cq}}$ of the mass octupole $h^{\mathrm{TT}}_{ij,\mathrm{oct}}$ and current quadrupole $h^{\mathrm{TT}}_{ij,\mathrm{cq}}$ is the next-to-leading order contribution. These are given by \cite{misner_1973, maggiore_vol1}
		\begin{align}
			h^{\mathrm{TT}}_{ij,\mathrm{quad}} &= \frac{2}{r} \Lambda_{ijk\ell} \qty[ \ddot{M}^{k\ell} ]_{t'}, \\
			h^{\mathrm{TT}}_{ij,\mathrm{oct}} &= \frac{2}{3r} \Lambda_{ijk\ell} n_{p} \qty[ \dddot{M}^{k\ell p} ]_{t'}, \\
			h^{\mathrm{TT}}_{ij,\mathrm{cq}} &= \frac{4}{3r} \Lambda_{ijk\ell} n_{p} \qty[ \dddot{P}^{k\ell p} + \dddot{P}^{\ell kp} - 2\dddot{P}^{pk\ell} ]_{t'}.
		\end{align}
		Here, $t'$ is the retarded time, and we have the multipole moments of the mass and momentum density:
		\begin{align}
			M^{k_1 k_2 \cdots k_{\ell}} &= \int \dd[3]{x} T^{00}(t, \vb{x}) x^{k_1} x^{k_2} \cdots x^{k_{\ell}}, \\
			P^{ik_1 k_2 \cdots k_{\ell}} &= \int \dd[3]{x} T^{0i}(t, \vb{x}) x^{k_1} x^{k_2} \cdots x^{k_\ell},
		\end{align}
		
		In the full PN formalism, the mass quadrupole corresponds to the 0PN waveform, while both the mass octupole and current quadrupole waveforms correspond to the 0.5PN waveforms. We can calculate both contributions to the waveform in the linear theory since corrections to the energy-momentum tensor of the source first appear at 1PN \cite{maggiore_vol1}.
		
		Starting from $\vu{n}$, we construct a right-handed orthonormal frame $\{ \vu{u}, \vu{v}, \vu{n} \}$. The two independent components of $h_{ab}$ in the TT gauge represent the two polarization states of the GW called the plus and cross modes:
		\begin{align}
			h^{+} &= \frac{1}{2} (u^i u^j - v^i v^j) h_{ij}^{\mathrm{TT}}, \\
			h^{\times} &= \frac{1}{2} (u^i v^j + u^j v^i) h_{ij}^{\mathrm{TT}}.
		\end{align}
		Far from the source, the (average) rate at which GWs remove energy from the system is \cite{poisson_2014}
		\begin{equation}
			\dv{E}{t} = \frac{r^2}{32\pi} \oint \dd{\Omega} \expval{\dot{h}_{ij}^{\mathrm{TT}} \dot{h}^{ij}_{\mathrm{TT}}},
		\end{equation}
		where $\expval{\cdot}$ is a temporal average over several periods. GWs also radiate away angular momentum from the system. For quasicircular orbits, the angular momentum flux is proportional to the energy flux \cite{blanchet_2024}. Thus, the angular momentum flux does not provide any additional constraint on the evolution of quasicircular systems.
        
	\section{Quasi-circular Binary Systems and Lagrange Three-body Systems}\label{sec:waveforms}
		\subsection{Waveforms of the binary system}
			Consider a binary system composed of two masses $m_{1,\mathrm{(2B)}}$ and $m_{2,\mathrm{(2B)}}$ separated by a distance $a$, and whose center-of-mass (COM) is located at a distance $r_{\mathrm{(2B)}}$ from the observer. Let $M_{\mathrm{(2B)}} = m_{1,\mathrm{(2B)}} + m_{2,\mathrm{(2B)}}$ be the total mass and $\alpha_i = m_{i,\mathrm{(2B)}}/M_{\mathrm{(2B)}}$, $i = 1,2$ be the normalized mass. Note that
			\begin{equation}
				\alpha_1 + \alpha_2 = 1.
			\end{equation}
			We work in the COM frame. In the Newtonian approximation, the orbital frequency of the binary system is
			\begin{equation}
				\omega_{\mathrm{(2B)}}^2 = \frac{M_{\mathrm{(2B)}}}{a^3}.
			\end{equation}
			As a simple model, there are five independent parameters describing the GW of a binary system: the initial separation distance $a_0 = a(0)$, masses $m_{1,\mathrm{(2B)}}$ and $m_{2,\mathrm{(2B)}}$ of the bodies, distance $r_{\mathrm{(2B)}}$ to the detector, and the orbital inclination angle $\iota_{\mathrm{(2B)}}$. For circular orbits, $a_0$ is the same as the semimajor axis of the elliptical orbit of the reduced mass. As it is usual for circular orbits, we set the argument of pericenter $\phi_{\mathrm{(2B)}}$ to zero. The plus and cross modes up to the 0.5PN approximation are \cite{maggiore_vol1, poisson_2014, creighton_2011}
			\begin{widetext}
				\begin{align}
					\begin{split}\label{eq:2B_totalGW_plus}
						&h_{\mathrm{(2B)}}^{+} = \mathcal{A}_{\mathrm{quad,(2B)}}^{+} \cos(2\Phi_{\mathrm{(2B)}}) + \qty(\mathcal{A}_{\mathrm{oct+cq,(2B)}}^{+})_{\omega} \cos(\Phi_{\mathrm{(2B)}}) - \qty(\mathcal{A}_{\mathrm{oct+cq,(2B)}}^{+})_{3\omega} \cos(3\Phi_{\mathrm{(2B)}}),
					\end{split}
					\\
					\begin{split}\label{eq:2B_totalGW_cross}
						&h_{\mathrm{(2B)}}^{\times} = \mathcal{A}_{\mathrm{quad,(2B)}}^{\times} \sin(2\Phi_{\mathrm{(2B)}}) + \qty(\mathcal{A}_{\mathrm{oct+cq,(2B)}}^{\times})_{\omega} \sin(\Phi_{\mathrm{(2B)}}) - \qty(\mathcal{A}_{\mathrm{oct+cq,(2B)}}^{\times})_{3\omega} \sin(3\Phi_{\mathrm{(2B)}}),
					\end{split}
					\\
					&\mathcal{A}_{\mathrm{quad,(2B)}}^{+} = \frac{4 M_{(\mathrm{2B})}^{5/3} \omega_{\mathrm{(2B)}}^{2/3}\alpha_1 \alpha_2 }{r_{(\mathrm{2B})}} \frac{[1 + \cos^2(\iota_{(\mathrm{2B})})]}{2}, \label{eq:2B_ampQuad_plus} \\
					&\mathcal{A}_{\mathrm{quad,(2B)}}^{\times} = \frac{4 M_{(\mathrm{2B})}^{5/3} \omega_{\mathrm{(2B)}}^{2/3} \alpha_1 \alpha_2 }{r_{(\mathrm{2B})}} \cos(\iota_{(\mathrm{2B})}), \label{eq:2B_ampQuad_cross} \\
					&\qty(\mathcal{A}_{\mathrm{oct+cq,(2B)}}^{+})_{\omega} = \frac{1}{4} \frac{ M_{(\mathrm{2B})}^2 \alpha_1 \alpha_2 (\alpha_2 - \alpha_1) \omega_{\mathrm{(2B)}} }{r_{(\mathrm{2B})}} \sin(\iota_{\mathrm{(2B)}}) \qty[5 + \cos[2](\iota_{(\mathrm{2B})})], \label{eq:2B_amp_w_plus} \\
					&\qty(\mathcal{A}_{\mathrm{oct+cq,(2B)}}^{\times})_{\omega} = \frac{3}{4} \frac{ M_{(\mathrm{2B})}^2 \alpha_1 \alpha_2 (\alpha_2 - \alpha_1) \omega_{\mathrm{(2B)}} }{r_{(\mathrm{2B})}} \sin(2\iota_{(\mathrm{2B})}), \label{eq:2B_amp_w_cross} \\
					&\qty(\mathcal{A}_{\mathrm{oct+cq,(2B)}}^{+})_{3\omega} = \frac{9}{4} \frac{ M_{(\mathrm{2B})}^2 \alpha_1 \alpha_2 (\alpha_2 - \alpha_1) \omega_{\mathrm{(2B)}} }{r_{(\mathrm{2B})}} \sin(\iota_{\mathrm{(2B)}}) \qty[1 + \cos[2](\iota_{(\mathrm{2B})})], \label{eq:2B_amp_3w_plus} \\
					&\qty(\mathcal{A}_{\mathrm{oct+cq,(2B)}}^{\times})_{3\omega} = \frac{9}{4} \frac{ M_{(\mathrm{2B})}^2 \alpha_1 \alpha_2 (\alpha_2 - \alpha_1) \omega_{\mathrm{(2B)}} }{r_{(\mathrm{2B})}} \sin(2\iota_{(\mathrm{2B})}). \label{eq:2B_amp_3w_cross}
				\end{align}
			\end{widetext}
			
			Assuming that the binary is evolving adiabatically so that it remains quasicircular, we have
			\begin{align}
				\dv{\omega_{\mathrm{(2B)}}}{t} &= \frac{96}{5} \mathcal{M}_{(\mathrm{2B})}^{5/3} \omega_{\mathrm{(2B)}}^{11/3}, \label{eq:2BorbitalFreqODE} \\
				a^3 \dv{a}{t} &= -\frac{64}{5} M_{(\mathrm{2B})}^{4/3} \mathcal{M}_{(\mathrm{2B})}^{5/3} \label{eq:2BsepDistanceODE},
			\end{align}
			where the chirp mass of the binary is
			\begin{equation}
				\mathcal{M}_{\mathrm{(2B)}} = M_{\mathrm{(2B)}} (\alpha_1 \alpha_2)^{3/5}.
			\end{equation}
			Generally then, once we consider radiative losses, the orbital frequency and hence the gravitational wave frequency become time dependent. The finite value at which $\omega_{(\mathrm{2B})}$ and $a$ diverge is the coalescence time $t_{\mathrm{c,(2B)}}$:
			\begin{equation}
				t_{\mathrm{c,(2B)}} = \frac{5}{256} \frac{a_0^4}{M_{(\mathrm{2B})}^{4/3} \mathcal{M}_{(\mathrm{2B})}^{5/3}}, \quad a_0 = a(0).
			\end{equation}
			For an adiabatic evolution, the waveform has the same form as that in \eqref{eq:2B_totalGW_plus} and \eqref{eq:2B_totalGW_cross}; the only difference is that factors of $\omega_{(\mathrm{2B})}$ appearing in the amplitude are now time dependent, and the phase is replaced by
			\begin{equation}
				\begin{split}
				\Phi_{\mathrm{(2B)}}(t) &= -\qty(\frac{t_{\mathrm{c,(2B)}} - t}{5\mathcal{M}_{\mathrm{(2B)}}})^{5/8} + \Phi_{\mathrm{c,(2B)}},
				\end{split}
			\end{equation}
			where $\Phi_{\mathrm{c,(2B)}}$ is an integration constant defined as the value of the phase $\Phi_{(\mathrm{2B})}$ at the coalescence time $t_{\mathrm{c,(2B)}}$. Note that the corrections to the phase appear starting only at 1PN, which is beyond the scope of this work.
			
		\subsection{Waveforms of the Lagrange three-body system}
\begin{figure*}[htb!]
				\centering
				{\makebox[0.45\textwidth]{{\includegraphics[width=0.45\textwidth]{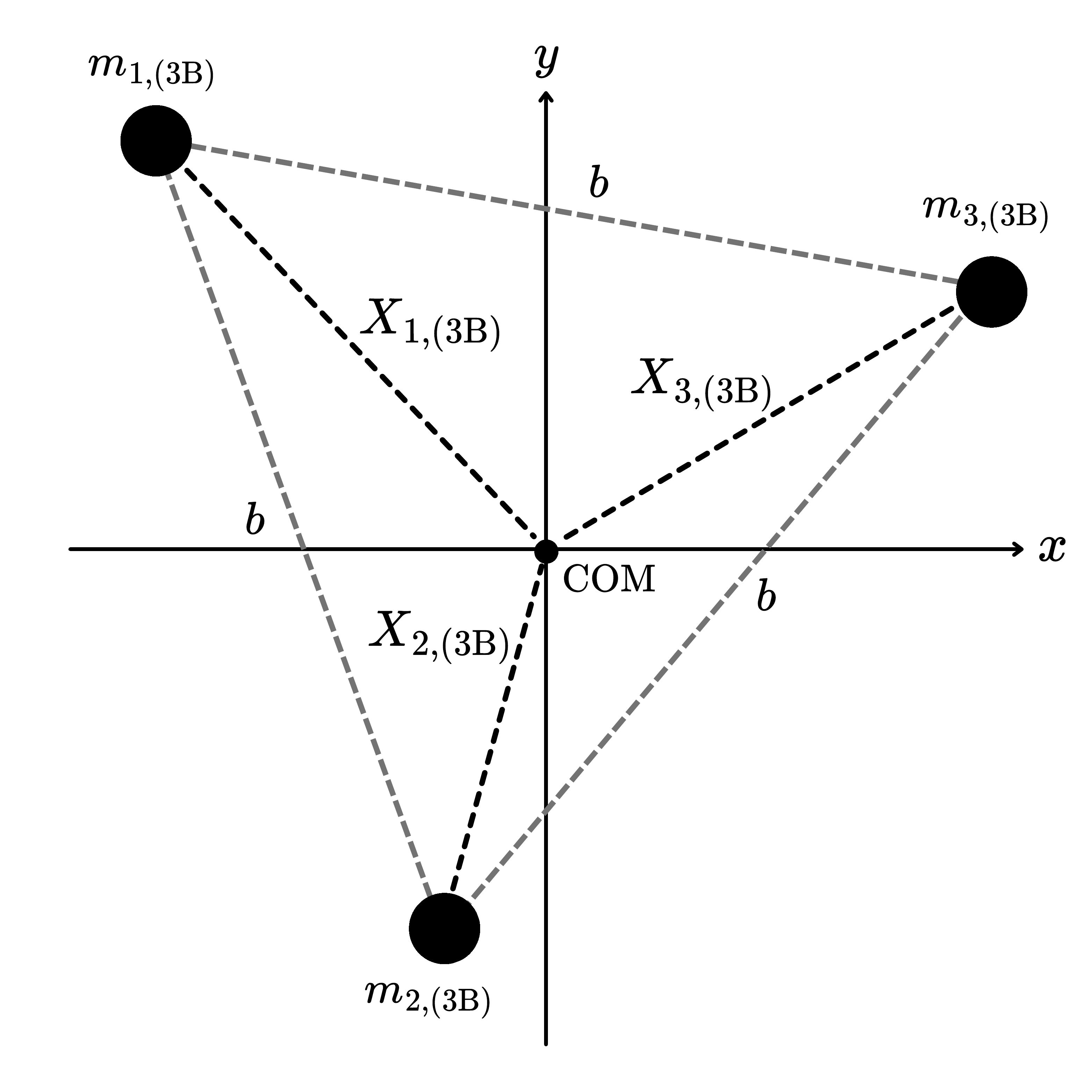}}}}
				\caption{Schematic diagram of the Lagrange three-body system. Each mass moves in a circular orbit about the COM.}
				\label{fig:lagrangeConfiguration}
			\end{figure*}

			Next, consider a Lagrange three-body system composed of three masses $m_{i,\mathrm{(2B)}}$, $i = 1,2,3$, each separated by a distance $b$ (Figure \ref{fig:lagrangeConfiguration}), and whose COM is located at a distance $r_{\mathrm{(3B)}}$ from the detector. Let $M_{\mathrm{(3B)}} = m_{1,\mathrm{(3B)}} + m_{2,\mathrm{(3B)}} + m_{3,\mathrm{(3B)}}$ be the total mass and $\beta_i = m_{i,\mathrm{(3B)}}/M_{\mathrm{(3B)}}$, $i = 1,2,3$ be the normalized mass. The orbital frequency of this system is
			\begin{equation}
				\omega_{\mathrm{(3B)}}^2 = \frac{M_{\mathrm{(3B)}}}{b^3}.
			\end{equation}
			Note also that
			\begin{equation}\label{eq:3B_massRatioSum}
				\beta_1 + \beta_2 + \beta_3 = 1.
			\end{equation}

			Because there are now three masses, we have six parameters describing the GW of the Lagrange triple: the initial separation distance $b_0 = b(0)$, three masses $m_{i,\mathrm{(3B)}}$, $i = 1,2,3$, distance $r_{\mathrm{(3B)}}$ to the detector, and the orbital inclination angle $\iota_{\mathrm{(3B)}}$. Following Asada \cite{asada_2009}, the plus and cross modes up to the 0.5PN approximation are
			\begin{widetext}
				\begin{align}
					\begin{split}\label{eq:3B_totalGW_plus}
						&h_{\mathrm{(3B)}}^{+} = \mathcal{A}_{\mathrm{quad,(3B)}}^{+} \cos(2\Phi_{\mathrm{(3B)}} + \Psi_{\mathrm{quad, (3B)}}) + \qty(\mathcal{A}_{\mathrm{oct+cq,(3B)}}^{+})_{\omega} \cos(\Phi_{\mathrm{(3B)}} + \Psi_{\mathrm{\omega,(3B)}}) \\
						&\qquad\qquad - \qty(\mathcal{A}_{\mathrm{oct+cq,(3B)}}^{+})_{3\omega} \cos(3\Phi_{\mathrm{(3B)}} + \Psi_{\mathrm{3\omega,(3B)}}),
					\end{split}
					\\
					\begin{split}\label{eq:3B_totalGW_cross}
						&h_{\mathrm{(3B)}}^{\times} = \mathcal{A}_{\mathrm{quad,(3B)}}^{\times} \sin(2\Phi_{\mathrm{(3B)}} + \Psi_{\mathrm{quad, (3B)}}) + \qty(\mathcal{A}_{\mathrm{oct+cq,(3B)}}^{\times})_{\omega} \sin(\Phi_{\mathrm{(3B)}} + \Psi_{\mathrm{\omega,(3B)}}) \\
						&\qquad\qquad - \qty(\mathcal{A}_{\mathrm{oct+cq,(3B)}}^{\times})_{3\omega} \sin(3\Phi_{\mathrm{(3B)}} + \Psi_{\mathrm{3\omega,(3B)}}),
					\end{split}
					\\
					\begin{split}\label{eq:3B_ampQuad_plus}
						&\mathcal{A}_{\mathrm{quad,(3B)}}^{+} = \frac{4 M_{(\mathrm{3B})}^{5/3} \omega_{\mathrm{(3B)}}^{2/3} \sqrt{(\beta_1\beta_2 + \beta_2\beta_3 + \beta_1\beta_3)^2 - 3(\beta_1^2\beta_2\beta_3 + \beta_1\beta_2^2\beta_3 + \beta_1\beta_2\beta_3^2)}}{r_{(\mathrm{3B})}} \frac{1 + \cos[2](\iota_{(\mathrm{3B})})}{2},
					\end{split}
					\\
					\begin{split}\label{eq:3B_ampQuad_cross}
						&\mathcal{A}_{\mathrm{quad,(3B)}}^{\times} = \frac{4 M_{(\mathrm{3B})}^{5/3} \omega_{\mathrm{(3B)}}^{2/3} \sqrt{(\beta_1\beta_2 + \beta_2\beta_3 + \beta_1\beta_3)^2 - 3(\beta_1^2\beta_2\beta_3 + \beta_1\beta_2^2\beta_3 + \beta_1\beta_2\beta_3^2)}}{r_{(\mathrm{3B})}} \cos(\iota_{(\mathrm{3B})}),
					\end{split}
					\\
					\begin{split}\label{eq:3B_amp_w_plus}
						&\qty(\mathcal{A}_{\mathrm{oct+cq,(3B)}}^{+})_{\omega} = \frac{1}{8} \frac{ M_{(\mathrm{3B})}^2 \omega_{\mathrm{(3B)}}}{r_{(\mathrm{3B})}} \sin(\iota_{(\mathrm{3B})}) [5 + \cos[2](\iota_{(\mathrm{3B})})] \\
						&\qquad\qquad\qquad\qquad \times \sqrt{ (\beta_1 - \beta_2)^2 [(\beta_3 - \beta_1)(\beta_3 - \beta_2) - 3\beta_1\beta_2]^2 + 3\beta_3^2 [\beta_1 (\beta_1 - \beta_3 ) + \beta_2 (\beta_2 - \beta_3)]^2 },
					\end{split}
					\\
					\begin{split}\label{eq:3B_amp_w_cross}
						&\qty(\mathcal{A}_{\mathrm{oct+cq,(3B)}}^{\times})_{\omega} =\frac{3}{8} \frac{ M_{(\mathrm{3B})}^2 \omega_{\mathrm{(3B)}}}{r_{(\mathrm{3B})}} \sin(2\iota_{(\mathrm{3B})}) \\
						&\qquad\qquad\qquad\qquad \times \sqrt{ (\beta_1 - \beta_2)^2 [(\beta_3 - \beta_1)(\beta_3 - \beta_2) - 3\beta_1\beta_2]^2 + 3\beta_3^2 [\beta_1 (\beta_1 - \beta_3 ) + \beta_2 (\beta_2 - \beta_3)]^2 },
					\end{split}
					\\
					\begin{split}\label{eq:3B_amp_3w_plus}
						&\qty(\mathcal{A}_{\mathrm{oct+cq,(3B)}}^{+})_{3\omega} = \frac{9}{4} \frac{M_{(\mathrm{3B})}^2 \omega_{\mathrm{(3B)}}}{r_{(\mathrm{3B})}} \sin(\iota_{(\mathrm{3B})}) [1 + \cos[2](\iota_{(\mathrm{3B})})] \sqrt{27\beta_1^2 \beta_2^2 \beta_3^2 + (\beta_1 - \beta_2)^2 (\beta_2 - \beta_3)^2 (\beta_3 - \beta_1)^2},
					\end{split}
					\\
					\begin{split}\label{eq:3B_amp_3w_cross}
						&\qty(\mathcal{A}_{\mathrm{oct+cq,(3B)}}^{\times})_{3\omega} = \frac{9}{4} \frac{M_{(\mathrm{3B})}^2 \omega_{\mathrm{(3B)}}}{r_{(\mathrm{3B})}} \sin(2\iota_{(\mathrm{3B})}) \sqrt{27\beta_1^2 \beta_2^2 \beta_3^2 + (\beta_1 - \beta_2)^2 (\beta_2 - \beta_3)^2 (\beta_3 - \beta_1)^2}.
					\end{split}
				\end{align}
			\end{widetext}
			
			Unlike the binary system, the 0.5PN waveform does not simply depend on the mass differences. Even when all the masses in the Lagrange triple are equal, the 0.5PN waveform does not vanish. Furthermore, the waveform of the Lagrange triple has phase shifts given by
			\begin{align}
				&\tan(\Psi_{\mathrm{quad,(3B)}}) = \frac{\sqrt{3} \beta_3 (\beta_2 - \beta_1)}{2\beta_1\beta_2 - \beta_3 (\beta_1 + \beta_2)}, \\
				&\tan(\Psi_{\mathrm{\omega,(3B)}}) = \frac{(\beta_1 - \beta_2)[3\beta_1 \beta_2 - (\beta_3-\beta_1)(\beta_3 - \beta_2)]}{\sqrt{3} \beta_3[\beta_1 (\beta_1 - \beta_3) + \beta_2 (\beta_2 - \beta_3)]}, \\
				&\tan(\Psi_{\mathrm{3\omega,(3B)}}) = \frac{(\beta_1 - \beta_2) (\beta_2 - \beta_3)(\beta_3 - \beta_1)}{3\sqrt{3} \beta_1 \beta_2 \beta_3 }.
			\end{align}
			These phase shifts give rise to time lags between the mass quadrupole waveform and the 0.5PN waveform \cite{asada_2009}:
			\begin{align}
				\Delta t_{\omega} &= \frac{\Psi_{\mathrm{\omega,(3B)}}}{\omega_{\mathrm{(3B)}}} - \frac{\Psi_{\mathrm{quad,(3B)}}}{2\omega_{\mathrm{(3B)}}}, \label{eq:3B_timeLags_omega} \\
				\Delta t_{3\omega} &= \frac{\Psi_{\mathrm{3\omega,(3B)}}}{3\omega_{\mathrm{(3B)}}} - \frac{\Psi_{\mathrm{quad,(3B)}}}{2\omega_{\mathrm{(3B)}}}, \label{eq:3B_timeLags_3omega}
			\end{align}
			Phase shifts that cannot be removed by a time translation are not unique to the Lagrange triple. An overall phase shift in the mass quadrupole GWs also exists for the circular restricted three-body system when the test mass sits at one of the triangular Lagrange points \cite{barandiaran_2024}.
			
			We argue that the phase shifts, and thus the time lags, arise from the asymmetry of the triangle configuration. On the one hand, in a binary system, the initial angular positions of the masses in the COM frame can always be chosen to be 0 and $\pi$. On the other hand, in the Lagrange three-body system, the COM can be located anywhere in the equilateral triangle. As a result, the initial angular positions of the masses can take on different values, making the initial orbital phase important for the waveform. The identifications $m_{1,\mathrm{(3B)}}$, $m_{2,\mathrm{(3B)}}$, and $m_{3,\mathrm{(3B)}}$ will only lead to the same waveform when the assignment is done under a cyclic permutation of $\qty{m_{1,\mathrm{(3B)}}, m_{2,\mathrm{(3B)}}, m_{3,\mathrm{(3B)}}}$.

            As the Lagrange triple emits GWs, there is a radiation reaction force on each body due to the mass quadrupole waveform. Yamada \textit{et al.} \cite{yamada_2016} have shown that during the early stages of the inspiral, the Lagrange three-body system shrinks adiabatically in a nonchaotic manner. Thus, we can assume an adiabatic evolution so that
			\begin{align}
				\dv{\omega_{\mathrm{(3B)}}}{t} &= \frac{96}{5} \mathcal{M}_{(\mathrm{3B})}^{5/3} \omega_{\mathrm{(3B)}}^{11/3}, \label{eq:3BorbitalFreqODE} \\
				b^3 \dv{b}{t} &= -\frac{64}{5} M_{(\mathrm{3B})}^{4/3} \mathcal{M}_{(\mathrm{3B})}^{5/3}, \label{eq:3BsepDistanceODE}
			\end{align}
			where the chirp mass $\mathcal{M}_{\mathrm{(3B)}}$ for the Lagrange triple is
			\begin{widetext}
				\begin{equation}
					\mathcal{M}_{(\mathrm{3B})} = M_{(\mathrm{3B})} \qty[ \frac{(\beta_1\beta_2 + \beta_2\beta_3 + \beta_1\beta_3)^2 - 3(\beta_1^2\beta_2\beta_3 + \beta_1\beta_2^2\beta_3 + \beta_1\beta_2\beta_3^2)}{\beta_1\beta_2 + \beta_1\beta_3 + \beta_2\beta_3} ]^{3/5} \label{eq:3BchirpMass}.
				\end{equation}
			\end{widetext}
			Note that this expression is equivalent to the definition of Torigoe \textit{et al.} \cite{torigoe_2009} and Asada \cite{asada_2009}. The waveform of the Lagrange triple with radiative losses has the same form as in \eqref{eq:3B_totalGW_plus} and \eqref{eq:3B_totalGW_cross} except that factors of $\omega_{\mathrm{(3B)}}$ in the amplitude become time dependent, and $\Phi_{\mathrm{(3B)}}$ is
			\begin{equation}
				\begin{split}
				\Phi_{\mathrm{(3B)}}(t) &= -\qty(\frac{t_{\mathrm{c,(3B)}} - t}{5\mathcal{M}_{\mathrm{(3B)}}})^{5/8} + \Phi_{\mathrm{c,(3B)}},
				\end{split}
			\end{equation}
			where $\Phi_{\mathrm{c,(3B)}}$ is an integration constant and $t_{\mathrm{c,(3B)}}$ is the coalescence time of the triple:
			\begin{equation}
				t_{\mathrm{c,(3B)}} = \frac{5}{256} \frac{b_0^4}{M_{(\mathrm{3B})}^{4/3} \mathcal{M}_{(\mathrm{3B})}^{5/3}}, \quad b_0 = b(0).
			\end{equation}

            \begin{figure}[htbp!]
				\centering
				{\makebox[0.45\textwidth]{{\includegraphics[width=0.45\textwidth]{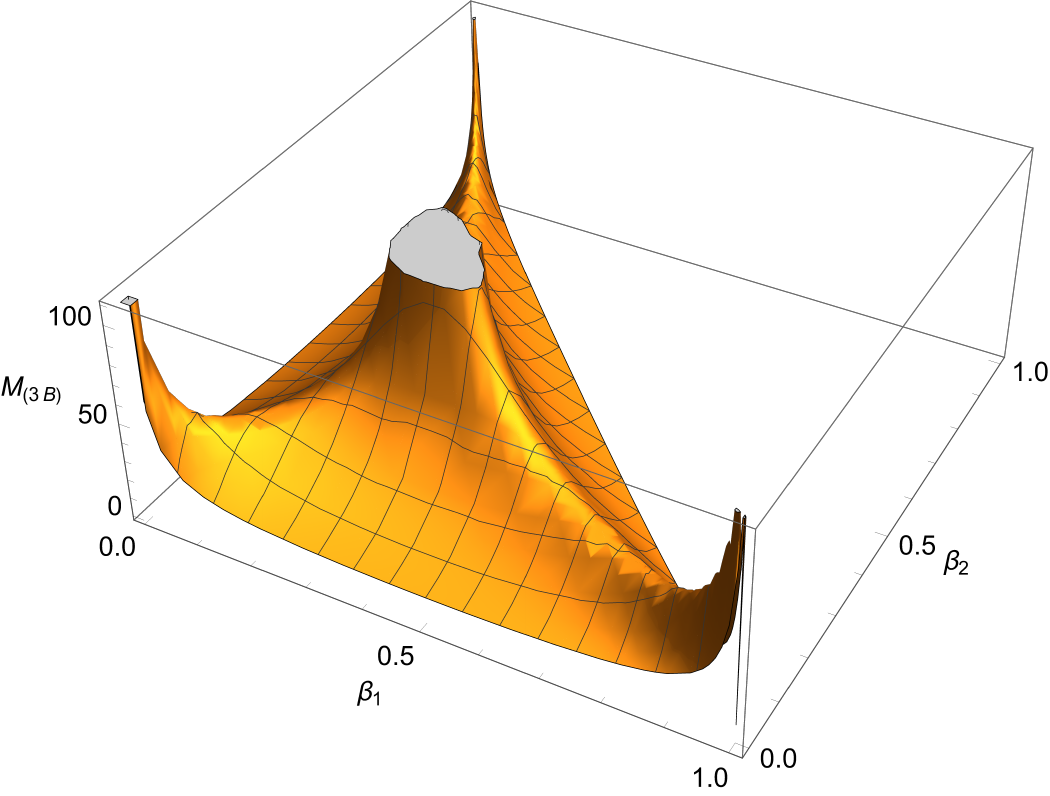}}}}
				\caption{Plot of the total mass $M_{\mathrm{(3B)}}$ in solar masses as a function of the normalized masses $\beta_1$ and $\beta_2$ for a chirp mass of $\mathcal{M}_{\mathrm{(3B)}} = 5 \ \mathrm{M}_{\odot}$.}
				\label{fig:totalMass}
			\end{figure}

            Note that the rate at which energy dissipates from each mass is generally different. This is because the orbital velocity of each mass depends on $X_{i,\mathrm{(3B)}}(t)$, and the energy of the individual masses depends on the velocity, so each mass dissipates energy at a different rate. However, what is relevant for the calculations is the dissipation of the \textit{total} energy of the system.
            
			This definition of the chirp mass was motivated by the fact that it must be the quantity that plays a direct role in the time evolution of the GW frequency, which can be calculated given only the waveform. Once we have the chirp mass, we can use equation \eqref{eq:3BchirpMass} to plot the total mass of the system as a function of two of the normalized masses; this is shown in Figure \ref{fig:totalMass}. Without additional restrictions to the normalized mass, the global minimum value of $M_{\mathrm{(3B)}}$ is the trivial value $M_{\mathrm{(3B)}} = 0$. However, if we were to observe astrophysical Lagrange three-body systems, we can impose the stability conditions to get a nonzero lower bound for the total mass $M_{\mathrm{(3B)}}$.

        \subsection{Stability of the Lagrange triple}
            In general, one can determine the stability of a mechanical system by identifying the equilibrium points, perturbing the equations of motion, and calculating the eigenvalues of the perturbed system evaluated at the equilibrium points. In Newtonian gravity, the Lagrange triple is stable if and only if \cite{gascheau_1843, sicardy_2010}
			\begin{equation}\label{eq:3B_stability1}
				\beta_1\beta_2 + \beta_2\beta_3 + \beta_1\beta_3 < \frac{1}{27}.
			\end{equation}
			The stability criterion for the Lagrange three-body system is generally strict. Linearly stable Lagrange three-body systems are composed of two very small masses and one large mass. At 1PN, the equilateral triangle configuration can only be realized when all the masses are equal \cite{yamada_2012}. For arbitrary masses, the triangular configuration is not always equilateral, and the stability criterion is stricter compared to the Newtonian one \cite{yamada_2015}.

            When it comes to the long-term stability, the radiation reaction force exerted on each mass will push the initially stable Lagrange triple to instability \cite{yamada_2016}. One of the bodies may become ejected, forming a hierarchical triple system or an isolated binary. However, to properly describe the stability of the Lagrange three-body system at long timescales \cite{yamada_2016}, one needs to incorporate strong field effects due to close encounters and high orbital velocities. Thus, as long as we are in the early stages of the inspiral and any environmental perturbations are small (e.g., in vacuum), there will be no strong field effects, and the stability of the Lagrange triple is dictated by \eqref{eq:3B_stability1}.
	
	\section{Waveform Degeneracy}\label{sec:degeneracy}
		\subsection{General method for finding degeneracies}
			Suppose we know all the parameters of the binary system. In this work, we determine the parameters of the Lagrange triple so that it has the same plus polarization waveform as a given binary. We choose $\{ b_0, M_{\mathrm{(3B)}}, \beta_1, \beta_2, r_{\mathrm{(3B)}}, \iota_{\mathrm{(3B)}} \}$ as the set of parameters describing the Lagrange triple instead of $\{ b_0, m_{1,\mathrm{(3B)}}, m_{2,\mathrm{(3B)}}, m_{3,\mathrm{(3B)}}, r_{\mathrm{(3B)}}, \iota_{\mathrm{(3B)}} \}$. The two sets are equivalent, since we can calculate the masses of the individual bodies if we know the total mass and use equation \eqref{eq:3B_massRatioSum}.
			
			First, we equate the amplitudes and phase of the waveforms of the binary and the Lagrange triple in the case where there are no radiative losses. Since the waveforms of a binary do not have observable time lags between the mass quadrupole waveform and 0.5PN waveform, waveform degeneracy occurs when $\Delta t_{\omega} = \Delta t_{3\omega} = 0$. This can be satisfied when
			\begin{equation}
			    \beta_1 = \beta_2.
			\end{equation}
			With this, it follows from equation \eqref{eq:3B_massRatioSum} that $\beta_1 = \beta_2 < 0.5$ and the number of independent parameters characterizing the Lagrange triple goes down to five: $\{ b_0, M_{\mathrm{(3B)}}, \beta_1, r_{\mathrm{(3B)}}, \iota_{\mathrm{(3B)}} \}$.
			
			Second, using the parameters we found in the case where there are no radiative losses (if there are any), the waveforms that include radiative losses can be identical at short timescales by choosing the phases $\Phi_{\mathrm{c,(2B)}}$ and $\Phi_{\mathrm{c,(3B)}}$ at coalescence. In this case, we then say that the binary and Lagrange triple are degenerate at short timescales. Note that $\Phi_{\mathrm{c,(2B)}}$ and $\Phi_{\mathrm{c,(3B)}}$ are not of astrophysical interest \cite{klein_2009}. For this reason, we arbitrarily set $\Phi_{\mathrm{c,(2B)}} = 0$ and choose $\Phi_{\mathrm{c,(3B)}} \in [0, 2\pi)$. We also check if it is possible for the waveforms of two systems to be completely identical up to the coalescence time. If this is the case, we say that there is a waveform degeneracy up to the coalescence time.
            
			Finally, after solving for the parameters of the Lagrange triple, we check if it is stable under linear perturbations. Since we will be concerned with Lagrange triples where $\beta_1 = \beta_2 < 0.5$. Then, the stability criterion translates to
			\begin{equation}\label{eq:3B_stability2}
				\beta_1 = \beta_2 < \frac{1}{9} (3 - 2\sqrt{2}) \approx 0.01906.
			\end{equation}
		
		\subsection{Degeneracy up to the mass quadrupole waveform}
			For mass quadrupole degeneracy at short timescales, we have
			\begin{align}
				\mathcal{A}_{\mathrm{quad,(2B)}}^{+} &= \mathcal{A}_{\mathrm{quad,(3B)}}^{+}, \label{eq:quadDegeneracy_matchQuad} \\
				\omega_{\mathrm{(2B)}}(0) &= \omega_{\mathrm{(3B)}}(0) = \sqrt{\frac{M_{\mathrm{(3B)}}}{b_0^3}}. \label{eq:quadDegeneracy_matchPhase}
			\end{align}
			Since we only have two equations for the five unknowns, we can choose three parameters of the Lagrange triple to be free parameters and then solve for the remaining two. We choose $\iota_{(\mathrm{3B})}$, $M_{\mathrm{(3B)}}$, and $\beta_1$ as the free parameters of the Lagrange triple. The remaining parameters, $r_{\mathrm{(3B)}}$ and $b_0$, must then satisfy
			\begin{align}
				r_{\mathrm{(3B)}} &= \qty( \frac{M_{\mathrm{(3B)}}}{\mathcal{M}_{\mathrm{(2B)}}} )^{5/3} \frac{r_{\mathrm{(2B)}} \beta_1 |3\beta_1 - 1| \qty[1 + \cos[2](\iota_{\mathrm{(3B)}})]}{\qty[1 + \cos[2](\iota_{\mathrm{(2B)}})]}, \label{eq:quadDegeneracy_r_3B_plus} \\
				b_0 &= \qty(\frac{M_{\mathrm{(3B)}}}{\qty(\omega_{\mathrm{(2B)}}(0))^{2}})^{1/3}. \label{eq:quadDegeneracy_b_plus}
			\end{align}
			The degeneracy between $h_{\mathrm{quad,(2B)}}^{\times}$ and $h_{\mathrm{quad,(3B)}}^{\times}$ proceeds along the same lines. The relevant amplitudes we need to equate are $\mathcal{A}_{\mathrm{quad,(2B)}}^{\times}$ and $\mathcal{A}_{\mathrm{quad,(3B)}}^{\times}$. By doing so, we obtain
			\begin{align}
				r_{\mathrm{(3B)}} &= \qty( \frac{M_{\mathrm{(3B)}}}{\mathcal{M}_{\mathrm{(2B)}}} )^{5/3} \frac{r_{\mathrm{(2B)}} \beta_1 |3\beta_1 - 1| \cos(\iota_{\mathrm{(3B)}})}{\cos(\iota_{\mathrm{(2B)}})}, \label{eq:quadDegeneracy_r_3B_cross} \\
				b_0 &= \qty(\frac{M_{\mathrm{(3B)}}}{\qty(\omega_{\mathrm{(2B)}}(0))^{2}})^{1/3}.  \label{eq:quadDegeneracy_b_cross}
			\end{align}
			Equations \eqref{eq:quadDegeneracy_r_3B_plus}--\eqref{eq:quadDegeneracy_b_plus} and \eqref{eq:quadDegeneracy_r_3B_cross}--\eqref{eq:quadDegeneracy_b_cross} only differ by their dependence on $\iota_{\mathrm{(3B)}}$. Thus, there can be degeneracy in the mass quadrupole waveform in both the plus and cross modes when
            \begin{equation}\label{eq:quadDegeneracy_i_3B_both}
                \iota_{\mathrm{(3B)}} = \iota_{\mathrm{(2B)}},
            \end{equation}
            not necessarily zero. In such a case, the distance $r_{\mathrm{(3B)}}$ of the Lagrange triple from the detector must be
			\begin{equation}\label{eq:quadDegeneracy_r_3B_both}
				r_{\mathrm{(3B)}} = \qty( \frac{M_{\mathrm{(3B)}}}{\mathcal{M}_{\mathrm{(2B)}}} )^{5/3} r_{\mathrm{(2B)}} \beta_1 |3\beta_1 - 1|.
			\end{equation}

            \begin{figure}[htbp!]
				\centering
				{\makebox[0.48\textwidth]{{\includegraphics[width=0.48\textwidth]{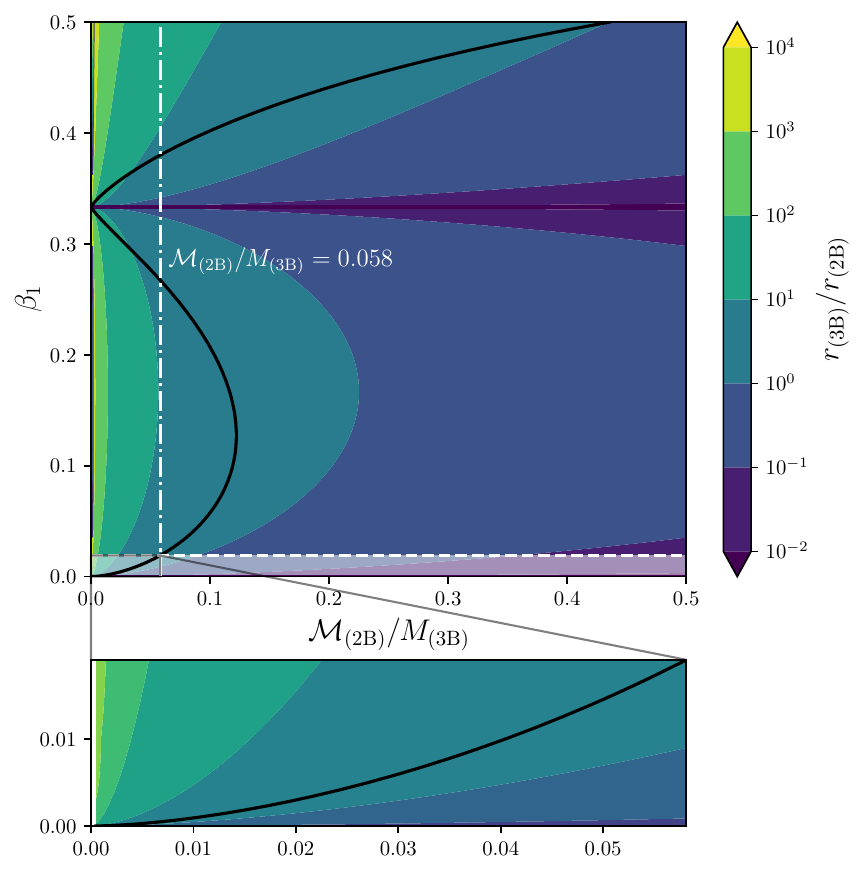}}}}
				\caption{Contour plot of $r_{\mathrm{(3B)}}/r_{\mathrm{(2B)}}$ in equation \eqref{eq:quadDegeneracy_r_3B_both} as a function of $\mathcal{M}_{\mathrm{(2B)}}/M_{\mathrm{(3B)}}$ and $\beta_1$. The translucent white region below the white dashed line is the Newtonian stability region of the Lagrange triple. The black curve correspond to values of $\beta_1$ satisfying equation \eqref{eq:equalChirpMass} for different values of $\mathcal{M}_{\mathrm{(2B)}}/M_{(\mathrm{3B})}$ and the white dash-dot line is approximately the last value of $\mathcal{M}_{\mathrm{(2B)}}/M_{(\mathrm{3B})}$ when $\beta_1$ falls within the Newtonian stability region. Note also that $\beta_1 < 0.5$ when $M_{\mathrm{(3B)}}/\mathcal{M}_{\mathrm{(2B)}} \lessapprox 0.435$. The lower plot enlarges the stability region.}
				\label{fig:contour_r_3B_both}
			\end{figure}
            
			Equations \eqref{eq:quadDegeneracy_b_cross}--\eqref{eq:quadDegeneracy_r_3B_both} give the parameters of the Lagrange triple for mass quadrupole degeneracy. Thus, the parameter space for mass quadrupole degeneracy is two-dimensional, parametrized by the total mass $M_{\mathrm{(3B)}}$ and the normalized mass $\beta_1$. From the contour plot of equation \eqref{eq:quadDegeneracy_r_3B_both} in Figure \ref{fig:contour_r_3B_both}, we see that linearly stable Lagrange triples that satisfy the mass quadrupole degeneracy are located at a comparable distance farther from the degenerate binary.

            Equations \eqref{eq:quadDegeneracy_b_cross}--\eqref{eq:quadDegeneracy_r_3B_both} also define a mapping from the space of binary systems to the space of Lagrange triples. However, such a mapping is not one to one in the sense that there can be multiple binaries that can correspond to the same Lagrange triple. This is because the binary is only specified by its initial orbital frequency, distance to the detector, and the orbital inclination angle.
            
            Now, in general, if a binary and a Lagrange triple are degenerate at short timescales, the system with the larger chirp mass coalesces first. Consequently, the two systems will have the same coalescence time when $\mathcal{M}_{\mathrm{(2B)}} = \mathcal{M}_{\mathrm{(3B)}}$ or equivalently,
			\begin{equation}\label{eq:equalChirpMass}
				\frac{\mathcal{M}_{\mathrm{(2B)}}}{M_{(\mathrm{3B})}} = \qty[ \frac{\beta_1 (1 - 3\beta_1)^2 }{2 - 3\beta_1} ]^{3/5}.
			\end{equation}
			
			The black points in Figure \ref{fig:contour_r_3B_both} are the values of $\beta_1$ satisfying $\mathcal{M}_{\mathrm{(2B)}} = \mathcal{M}_{\mathrm{(3B)}}$. When $\mathcal{M}_{\mathrm{(2B)}}/M_{(\mathrm{3B})} \lessapprox 0.058$, there are values of $\beta_1$ within the stability region. If we pick a value of $\beta_1$ within such a region, a stable Lagrange triple and a binary can be degenerate in the mass quadrupole waveform up to the coalescence time. Moreover, the chirp mass of the two systems can only be equal when $\mathcal{M}_{\mathrm{(2B)}}/M_{(\mathrm{3B})} \lessapprox 0.435$.

            \begin{figure*}[htbp!]
				\centering
				{\makebox[0.8\textwidth]{{\includegraphics[width=1\textwidth]{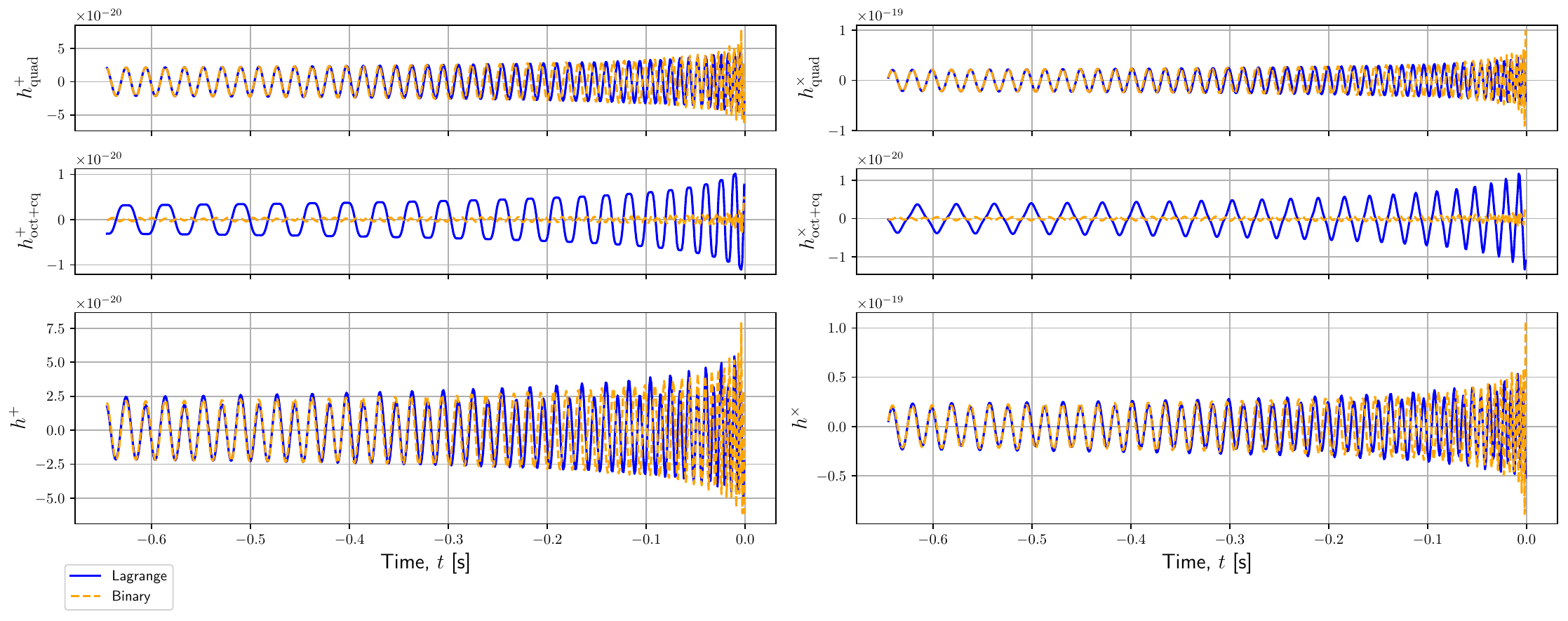}}}}
				\caption{Waveforms of \texttt{SymmetricBinary} and a linearly stable Lagrange triple ($b_0 = 3.0411 \times 10^{-11} \ \mathrm{pc}$, $m_{1,\mathrm{(3B)}} = m_{2,\mathrm{(3B)}} = 2.25 \ \mathrm{M}_{\odot}$, $m_{3,\mathrm{(3B)}} = 145.5 \ \mathrm{M}_{\odot}$, $r_{\mathrm{(3B)}} = 3.9744 \times 10^{6} \ \mathrm{pc}$, $\iota_{\mathrm{(3B)}} = 30^{\circ}$). Here, the binary coalesces first, and the matches up to the coalescence time are $M\qty(h^{+}_{\mathrm{(2B)}}, h^{+}_{\mathrm{(3B)}}) = M\qty(h^{\times}_{\mathrm{(2B)}}, h^{\times}_{\mathrm{(3B)}}) = 0.838$.}
				\label{fig:quadDegeneracy1}
			\end{figure*}

            \begin{figure*}[htbp!]
				\centering
				{\makebox[0.8\textwidth]{{\includegraphics[width=1\textwidth]{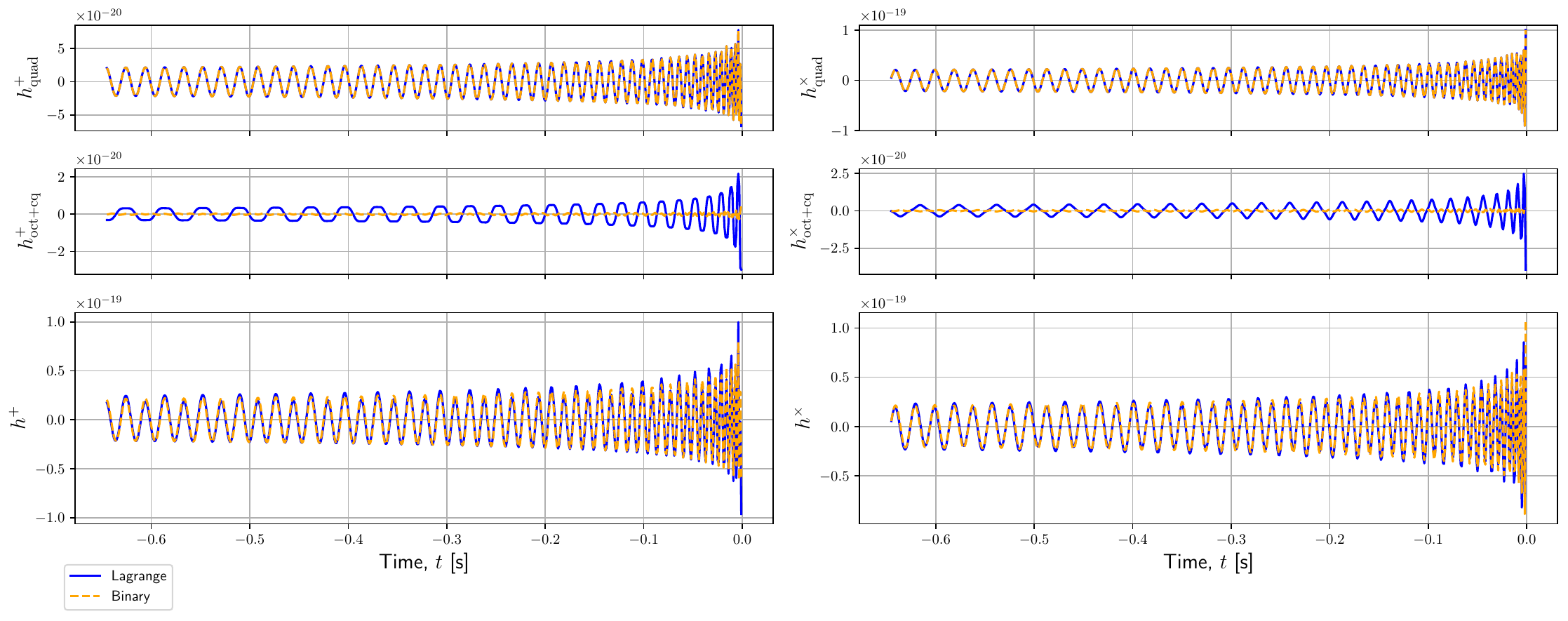}}}}
				\caption{Waveforms of \texttt{SymmetricBinary} and a linearly stable Lagrange triple ($b_0 = 3.0411 \times 10^{-11} \ \mathrm{pc}$, $m_{1,\mathrm{(3B)}} = m_{2,\mathrm{(3B)}} = 2.323 \ \mathrm{M}_{\odot}$, $m_{3,\mathrm{(3B)}} = 145.35 \ \mathrm{M}_{\odot}$, $r_{\mathrm{(3B)}} = 4.5260 \times 10^{6} \ \mathrm{pc}$, $\iota_{\mathrm{(3B)}} = 30^{\circ}$). Here, both systems have the same coalescence time, and the waveform matches up to the coalescence time are $M\qty(h^{+}_{\mathrm{(2B)}}, h^{+}_{\mathrm{(3B)}}) = 0.994$ and $M\qty(h^{\times}_{\mathrm{(2B)}}, h^{\times}_{\mathrm{(3B)}}) = 0.995$.}
				\label{fig:quadDegeneracy2}
			\end{figure*}

            \begin{figure*}[ht!]
				\centering
				{\makebox[0.7\textwidth]{{\includegraphics[width=0.75\textwidth]{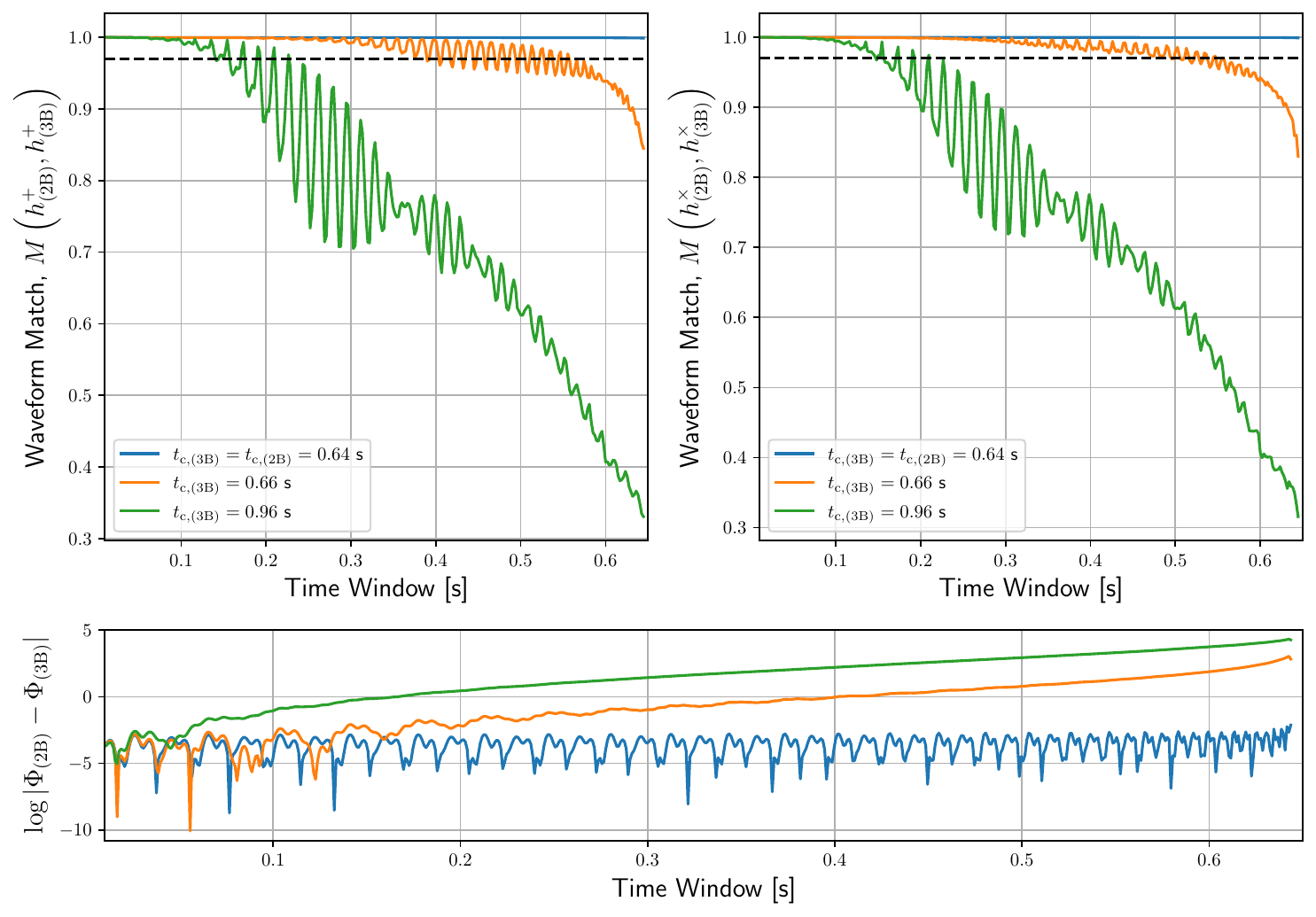}}}}
				\caption{Waveform match (top panels) and phase difference (bottom panel) of the plus and cross waveforms of \texttt{SymmetricBinary} and stable Lagrange triples with $\iota_{\mathrm{(3B)}} = \iota_{\mathrm{(2B)}} = 30^{\circ}$ and different coalescence times. The parameters of the Lagrange triple for each case are chosen in accordance with equations \eqref{eq:quadDegeneracy_b_cross}--\eqref{eq:quadDegeneracy_r_3B_both}. The black dashed lines in the top panels mark a waveform match of 0.97.}
				\label{fig:matchTimeSeries}
			\end{figure*}

            To illustrate our results, we consider a binary system with parameters $a_0 = 1.5 \times 10^{-11} \ \mathrm{pc}$, $m_{1,\mathrm{(2B)}} = 10 \ \mathrm{M}_{\odot}$, $m_{2,\mathrm{(2B)}} = 8 \ \mathrm{M}_{\odot}$, $r_{\mathrm{(2B)}} = 2 \times 10^{6} \ \mathrm{pc}$, and $\iota_{\mathrm{(2B)}} = 30^{\circ}$, hereafter referred to as \texttt{SymmetricBinary}. For the Lagrange triple, we choose $\beta_1 = 0.015$ and $M_{\mathrm{(3B)}} = 150 \ \mathrm{M}_{\odot}$; the smallness of $\beta_1$ ensures that the resulting Lagrange triple is linearly stable. Figure \ref{fig:quadDegeneracy1} shows an example of mass quadrupole degeneracy at short timescales, while Figure \ref{fig:quadDegeneracy2} shows an example of mass quadrupole degeneracy up to the coalescence time.
			
			Even if only the mass quadrupole waveforms are degenerate, the total waveforms of the two systems still look similar at short times. This is because the masses of the binary are comparable, making the octupolar contribution small. We can compare the similarity of two waveforms using the waveform match, which is the normalized inner product of the waveforms maximized over time and phase shifts \cite{owen_1996}:
			\begin{align}
				M(h_1, h_2) &= \max_{\Delta t, \Delta \phi} \frac{\bra{h_1}\ket{h_2}}{\norm{h_1} \norm{h_2}}, \\
				\bra{h_1}\ket{h_2} &= 4 \Re\qty[ \int_{0}^{+\infty}  \dd{f} \frac{\tilde{h}_1^*(f) \tilde{h}_2(f)}{S_n(f)} ].
			\end{align}
			Here, $\tilde{h}(f)$ denotes the Fourier transform of the waveform $h(t)$ and $S_n(f)$ is the power spectral density (PSD) of a detector. With the aid of \texttt{PyCBC} \cite{gwastropycbc_2024}, we use the simulated PSD of the advanced LIGO design generated from the LSC Algorithm Simulation Library \cite{lalsuite, abbott_2009} to calculate the waveform match of \texttt{SymmetricBinary} and stable Lagrange triples with different coalescence times. This is shown in Figure \ref{fig:matchTimeSeries}. As expected, the waveform match eventually decreases when the coalescence time of the two systems differs. Additionally, the oscillations in the waveform match come from the fact that at some points, the waveforms are identical, but eventually slide out of phase since they have different frequency sweeps.
			
			We highlight that the waveform match when $t_{\mathrm{c,(3B)}} = t_{\mathrm{c,(2B)}}$ remains above 0.97 even up to the coalescence time. Even in the case when $t_{\mathrm{c,(3B)}}$ only differs slightly from $t_{\mathrm{c,(2B)}}$, as shown by the orange line in Figure \ref{fig:matchTimeSeries}, the waveform match can remain above 0.97 for several cycles. Since the matched filtering procedure is also sensitive to phase variations, it is also important to look at the evolution of phase differences. As shown in the bottom panel of Figure \ref{fig:matchTimeSeries}, the phase difference reaches $10^0$ earlier when there is a larger difference between the coalescence times of the degenerate systems.
            
            Of course, the linearized theory of gravity and the PN theory fail well before the coalescence time. Nonetheless, this result still has important implications for GW data analysis during the inspiral phase since a waveform match above the standard threshold of 0.97 means that the detector is sensitive to at least 90\% of the signals in the parameter space \cite{owen_1996}.
		
		\subsection{Degeneracy up to the 0.5PN waveform}
			Although the mass quadrupole waveform is the dominant waveform, there are observed binary coalescences like GW190412 \cite{GW190412_2020} and GW190814 \cite{GW190814_2020} where the higher multipoles also make significant contributions to the total waveform due to asymmetric masses \cite{mahapatra_2024}. Additionally, the 0.5PN waveforms play a role in the parameter determination procedure studied by Asada \cite{asada_2009}. This leads us to consider the degeneracy up to the 0.5PN waveform. To this end, we require that
			\begin{align}
				\mathcal{A}_{\mathrm{quad,(2B)}}^{+} &= \mathcal{A}_{\mathrm{quad,(3B)}}^{+}, \label{eq:octDegeneracy_matchQuad_plus} \\
				\qty(\mathcal{A}_{\mathrm{oct+cq,(2B)}}^{+})_{\omega} &= \qty(\mathcal{A}_{\mathrm{oct+cq,(3B)}}^{+})_{\omega}, \label{eq:octDegeneracy_match_omega_plus} \\
				\qty(\mathcal{A}_{\mathrm{oct+cq,(2B)}}^{+})_{3\omega} &= \qty(\mathcal{A}_{\mathrm{oct+cq,(3B)}}^{+})_{3\omega}, \label{eq:octDegeneracy_match_3omega_plus} \\
				\omega_{\mathrm{(2B)}}(0) &= \omega_{\mathrm{(3B)}}(0) =  \sqrt{\frac{M_{\mathrm{(3B)}}}{b_0^3}}. \label{eq:octDegeneracy_matchPhase_plus}
			\end{align}
			Even up to the 0.5PN order, we only have four equations for five unknowns, hinting at a waveform degeneracy parametrized by one parameter. To determine the nature of this degeneracy, we begin by combining equations \eqref{eq:octDegeneracy_match_omega_plus} and \eqref{eq:octDegeneracy_match_3omega_plus}, to get
			\begin{align}
				\frac{\qty(\mathcal{A}_{\mathrm{oct+cq,(2B)}}^{+})_{\omega}}{\qty(\mathcal{A}_{\mathrm{oct+cq,(2B)}}^{+})_{3\omega}} &= \frac{ 5 + \cos[2](\iota_{(\mathrm{3B})}) }{ 18[1 + \cos[2](\iota_{(\mathrm{3B})})] } F(\beta_1), \label{eq:octDegeneracy_eqnForIota} \\
				F(\beta_1) &\coloneqq \frac{2}{3} \frac{|1 - 3\beta_1|}{\beta_1}.
			\end{align}
			Looking back at equations \eqref{eq:2B_amp_w_plus} and \eqref{eq:2B_amp_3w_plus}, note that
			\begin{equation}\label{eq:octDegeneracy_boundsForOctRatio}
				\frac{3}{9} \le \frac{\qty(\mathcal{A}_{\mathrm{oct+cq,(2B)}}^{+})_{\omega}}{\qty(\mathcal{A}_{\mathrm{oct+cq,(2B)}}^{+})_{3\omega}} \le \frac{5}{9}
			\end{equation}
			We can solve for $\iota_{(\mathrm{3B})}$ in equation \eqref{eq:octDegeneracy_eqnForIota}:
			\begin{equation}\label{eq:octDegeneracy_i_3B_plus}
				\sin(\iota_{(\mathrm{3B})}) = \sqrt{2 + \qty[\frac{1}{4} - \frac{27}{4} \frac{\qty(\mathcal{A}_{\mathrm{oct+cq,(2B)}}^{+})_{\omega}}{\qty(\mathcal{A}_{\mathrm{oct+cq,(2B)}}^{+})_{3\omega}} \frac{1}{F(\beta_1)}]^{-1} }.
			\end{equation}
			This motivates us to choose $\beta_1$ as the parameter characterizing the 0.5PN degeneracy in the plus mode. Next, we can find $b_0$, $M_{\mathrm{(3B)}}$, and $r_{\mathrm{(3B)}}$ by directly solving equations \eqref{eq:octDegeneracy_matchQuad_plus}--\eqref{eq:octDegeneracy_matchPhase_plus}. The results are
			
			\begin{widetext}
				\begin{align}
					\begin{split}\label{eq:octDegeneracy_M_3B_plus}
						M_{\mathrm{(3B)}} &= \frac{1}{2} \Biggl\{ \frac{\sqrt{3}(1 - 2\beta_1)}{8} \frac{\mathcal{A}^{+}_{\mathrm{quad,(2B)}}}{\qty(\mathcal{A}^{+}_{\mathrm{oct+cq,(2B)}})_{\omega}} \frac{\sin(\iota_{\mathrm{(3B)}})[5 + \cos[2](\iota_{\mathrm{(3B)}})]}{1 + \cos[2](\iota_{\mathrm{(3B)}})} \qty(\omega_{\mathrm{(2B)}}(0))^{1/3} \Biggr\}^{-3} + \frac{1}{2} \Biggl\{ \frac{9\sqrt{27}}{8} \\
						&\qquad \times \frac{\mathcal{A}^{+}_{\mathrm{quad,(2B)}}}{\qty(\mathcal{A}^{+}_{\mathrm{oct+cq,(2B)}})_{3\omega}} \frac{\beta_1 (1 - 2\beta_1)}{|3\beta_1 - 1|} \sin(\iota_{\mathrm{(3B)}}) \qty(\omega_{\mathrm{(2B)}}(0))^{1/3} \Biggr\}^{-3},
					\end{split}
					\\
					\begin{split}\label{eq:octDegeneracy_r_3B_plus}
						r_{(\mathrm{3B})} &= \frac{1}{\mathcal{A}^{+}_{\mathrm{quad,(2B)}} + \qty(\mathcal{A}^{+}_{\mathrm{oct+cq,(2B)}})_{\omega} + \qty(\mathcal{A}^{+}_{\mathrm{oct+cq,(2B)}})_{3\omega}} \Bigl\{  2M_{(\mathrm{3B})}^{5/3} \qty(\omega_{\mathrm{(2B)}}(0))^{2/3} \beta_1 |3\beta_1 - 1| [1 + \cos[2](\iota_{\mathrm{(3B)}})] \\
						&\quad + \frac{\sqrt{3}}{4} M_{(\mathrm{3B})}^2 \qty(\omega_{\mathrm{(2B)}}(0)) \beta_1 (1 - 2\beta_1) |3\beta_1 - 1| \sin(\iota_{(\mathrm{3B})}) [5 + \cos[2](\iota_{(\mathrm{3B})})] + \frac{9\sqrt{27}}{4} M_{(\mathrm{3B})}^2 \qty(\omega_{\mathrm{(2B)}}(0)) \\
						&\quad \times \beta_1^2 (1 - 2\beta_1) \sin(\iota_{(\mathrm{3B})}) [1 + \cos[2](\iota_{(\mathrm{3B})})] \Bigr\},
					\end{split}
					\\
					b_0 &= \qty(\frac{M_{\mathrm{(3B)}}}{\qty(\omega_{\mathrm{(2B)}}(0))^{2}})^{1/3}. \label{eq:octDegeneracy_b_plus}
				\end{align}
			\end{widetext}
			Note that because of the domain of the inverse sine function and equation \eqref{eq:octDegeneracy_boundsForOctRatio}, $\beta_1$ is bounded between
			\begin{align}
				\min(\beta_1) &= \qty[3 + \frac{9\qty(\mathcal{A}_{\mathrm{oct+cq,(2B)}}^{+})_{\omega}}{\qty(\mathcal{A}_{\mathrm{oct+cq,(2B)}}^{+})_{3\omega}}]^{-1}, \label{eq:octDegeneracy_minBeta} \\
				\max(\beta_1) &= 5 \qty[15 + \frac{27\qty(\mathcal{A}_{\mathrm{oct+cq,(2B)}}^{+})_{\omega}}{\qty(\mathcal{A}_{\mathrm{oct+cq,(2B)}}^{+})_{3\omega}}]^{-1}. \label{eq:octDegeneracy_maxBeta}
			\end{align}
			
			Since $\sin(\iota_{\mathrm{(3B)}})$ scales as $\beta_1^{-1/2}$, then in general, $M_{\mathrm{(3B)}}$ and $r_{\mathrm{(3B)}}$ scale as $\beta_1^{-3/2}$. Equality of the coalescence time and frequency sweep of the degenerate systems at short timescales is still dictated by $\mathcal{M}_{\mathrm{(2B)}} = \mathcal{M}_{\mathrm{(3B)}}$. However, unlike in the previous section where $M_{\mathrm{(3B)}}$ is a free parameter, $M_{\mathrm{(3B)}}$ now depends on $\beta_1$. This makes equation \eqref{eq:equalChirpMass} a more complicated equation for $\beta_1$, but it can still be solved numerically.
			
			Degeneracy in the cross modes proceeds along the same lines. Specifically, we require that
			\begin{align}
				\mathcal{A}_{\mathrm{quad,(2B)}}^{\times} &= \mathcal{A}_{\mathrm{quad,(3B)}}^{\times}, \label{eq:octDegeneracy_matchQuad_cross} \\
				\qty(\mathcal{A}_{\mathrm{oct+cq,(2B)}}^{\times})_{\omega} &= \qty(\mathcal{A}_{\mathrm{oct+cq,(3B)}}^{\times})_{\omega}, \label{eq:octDegeneracy_match_omega_cross} \\
				\qty(\mathcal{A}_{\mathrm{oct+cq,(2B)}}^{\times})_{3\omega} &= \qty(\mathcal{A}_{\mathrm{oct+cq,(3B)}}^{\times})_{3\omega}, \label{eq:octDegeneracy_match_3omega_cross}
			\end{align}
            along with equation \eqref{eq:octDegeneracy_matchPhase_plus}. For this set of nonlinear equations, it is difficult to isolate $r_{\mathrm{(3B)}}$, $M_{\mathrm{(3B)}}$, and $\iota_{\mathrm{(3B)}}$ in terms of $\beta_1$ and the parameters of the binary. Nonetheless, we can still numerically solve equations \eqref{eq:octDegeneracy_matchQuad_cross}--\eqref{eq:octDegeneracy_match_3omega_cross}.

            \begin{figure*}[htbp!]
				\centering
				{\makebox[1\textwidth]{{\includegraphics[width=1\textwidth]{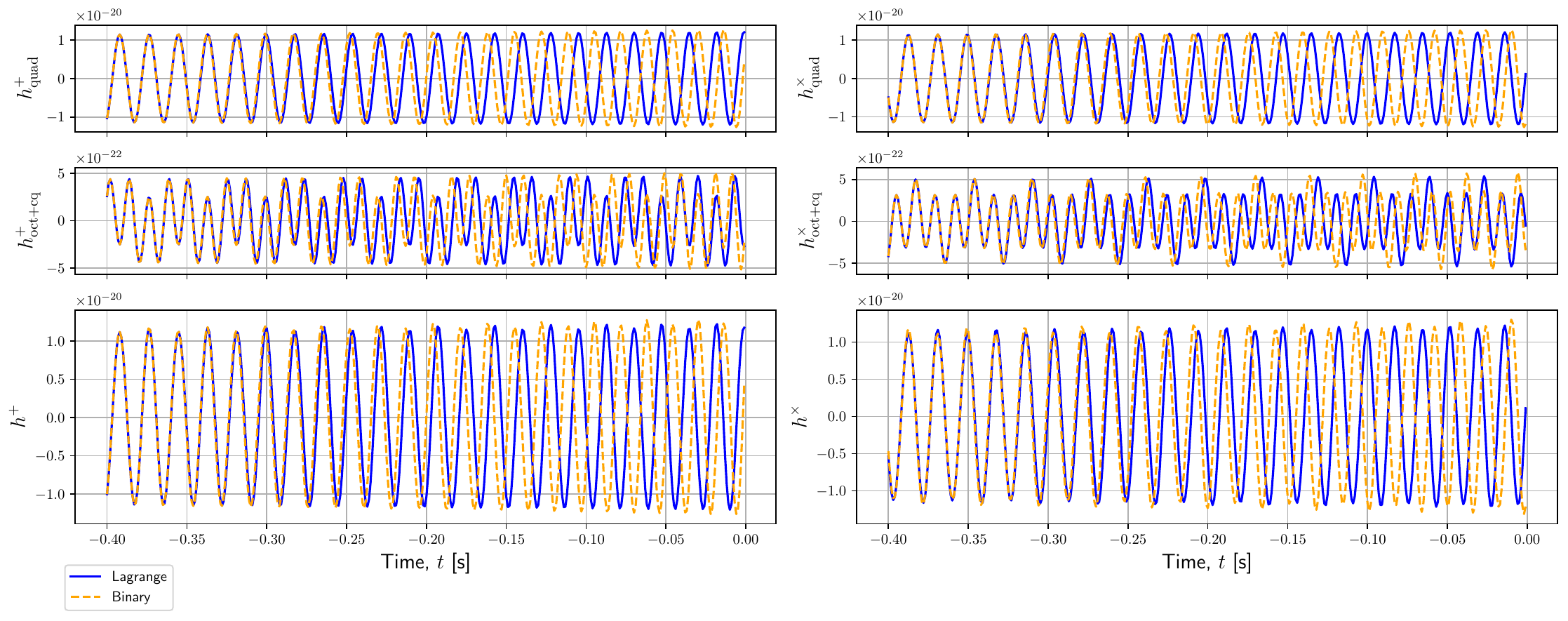}}}}
				\caption{Waveforms of \texttt{AsymmetricBinary} and a Lagrange three-body system ($b_0 = 1.6641 \times 10^{-11} \ \mathrm{pc}$, $m_{1,\mathrm{(3B)}} = m_{2,\mathrm{(3B)}} = 4.8965 \ \mathrm{M}_{\odot}$, $m_{3,\mathrm{(3B)}} = 19.5861 \ \mathrm{M}_{\odot}$, $r_{\mathrm{(3B)}} = 3.4578 \times 10^6 \ \mathrm{pc}$, $\iota_{\mathrm{(3B)}} = 5^{\circ}$). Here, the binary coalesces first, but the coalescence times of both systems are comparable ($t_{\mathrm{c,(2B)}} = 1.1505$ s, $t_{\mathrm{c,(3B)}} = 1.9963$ s). The waveform matches $M\qty(h^{+}_{\mathrm{(2B)}}, h^{+}_{\mathrm{(3B)}})$ and $M\qty(h^{\times}_{\mathrm{(2B)}}, h^{\times}_{\mathrm{(3B)}})$ in the specified time window are both around 0.22. This low value is because the dephasing of the two waveforms occurs early due to the differences in the coalescence times.}
				\label{fig:octDegeneracy1}
			\end{figure*}
            
            Ultimately, what we need is degeneracy in both polarization modes---that is, the parameter values such that the 0.5PN waveforms are degenerate in both polarization modes, making the systems indistinguishable regardless of the polarization mode detected. To obtain such parameter values, we begin by observing that equations \eqref{eq:octDegeneracy_match_omega_cross} and \eqref{eq:octDegeneracy_match_3omega_cross} satisfy
			\begin{equation}
				\frac{\qty(\mathcal{A}^{\times}_{\mathrm{oct+cq,(2B)}})_{\omega}}{\qty(\mathcal{A}^{\times}_{\mathrm{oct+cq,(2B)}})_{3\omega}} = \frac{1}{3} = \frac{1}{9} \frac{|\beta_1 - \beta_3|}{\beta_1}.
			\end{equation}
			Using equation \eqref{eq:3B_massRatioSum}, we get $\beta_1 = 1/6$. This value will always satisfy equations \eqref{eq:octDegeneracy_match_omega_cross} and \eqref{eq:octDegeneracy_match_3omega_cross} regardless of the value of $r_{\mathrm{(3B)}}$ and $M_{\mathrm{(3B)}}$, and as long as $\sin(2\iota_{(\mathrm{3B})}) \neq 0$.  
            
            Therefore, $\beta_1 = 1/6$, along with equations \eqref{eq:octDegeneracy_i_3B_plus}--\eqref{eq:octDegeneracy_b_plus}, define a mapping from the set of binary systems to the set of Lagrange triples so that there is 0.5PN degeneracy in both polarization modes. Interestingly, $\beta_1 = 1/6$ makes the Lagrange triple have the same orbital inclination angle as that of the binary ($\iota_{\mathrm{(2B)}} = \iota_{\mathrm{(3B)}}$). Unlike in the mass quadrupole degeneracy, the mapping for the 0.5PN degeneracy is one-to-one. To see why this is the case, note that in the 0.5PN degeneracy, we need to specify the binary via its initial orbital frequency and the amplitude of the waveforms up to the 0.5PN. We also note that $\iota_{\mathrm{(2B)}} = \iota_{\mathrm{(3B)}}$. This essentially amounts to five equations for the five parameters of the binary in our simplified model.

            \begin{figure}[hbp!]
				\centering
				\makebox[0.48\textwidth]{\includegraphics[width=0.48\textwidth]{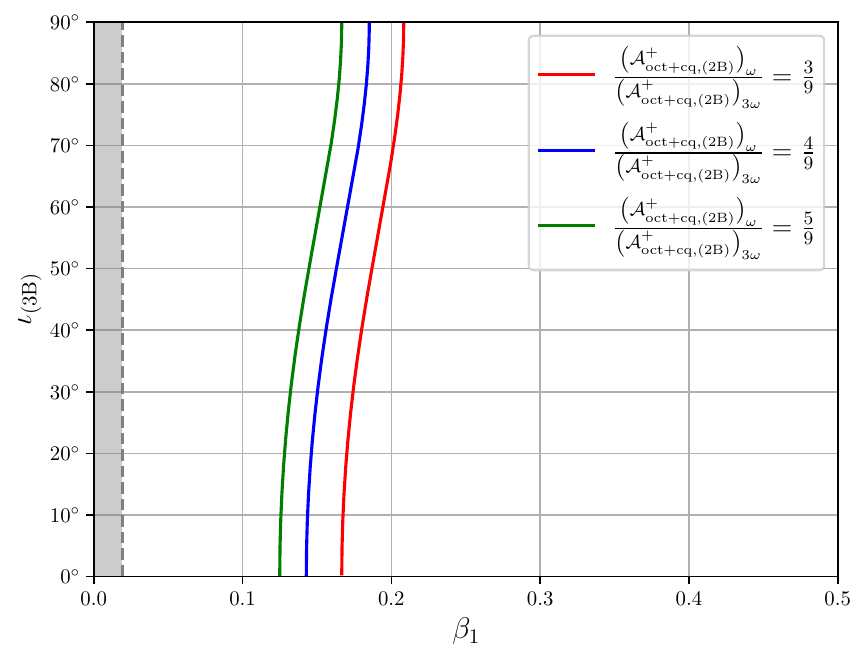}}
				\caption{Plot of \eqref{eq:octDegeneracy_i_3B_plus} for different values of $\qty(\mathcal{A}_{\mathrm{oct+cq,(2B)}}^{+})_{\omega}/\qty(\mathcal{A}_{\mathrm{oct+cq,(2B)}}^{+})_{3\omega}$. The gray region is the Newtonian stability region of a Lagrange triple. Observe that for any value of $3/9 \le \qty(\mathcal{A}_{\mathrm{oct+cq,(2B)}}^{+})_{\omega}/\qty(\mathcal{A}_{\mathrm{oct+cq,(2B)}}^{+})_{3\omega} \le 5/9$, the plot of $\iota_{\mathrm{(3B)}}$ does not fall within the Newtonian stability region.}
				\label{fig:iotaVsMassRatios}
			\end{figure}

            As the 0.5PN waveforms play an important role in binaries with asymmetric masses, the test binary we consider in this subsection has parameters $a_0 = 2.0 \times 10^{-11} \ \mathrm{pc}$, $m_{1,\mathrm{(2B)}} = 50 \ \mathrm{M}_{\odot}$, $m_{2,\mathrm{(2B)}} = 2 \ \mathrm{M}_{\odot}$, $r_{\mathrm{(2B)}} = 2 \times 10^{6} \ \mathrm{pc}$, and $\iota_{\mathrm{(2B)}} = 5^{\circ}$, hereafter referred to as \texttt{AsymmetricBinary}. To illustrate, we plot the degenerate waveforms in Figure \ref{fig:octDegeneracy1}. Note that the value $\beta_1 = 1/6$ is always within the allowed values for degeneracy, as dictated by equations \eqref{eq:octDegeneracy_minBeta}--\eqref{eq:octDegeneracy_maxBeta}. On the one hand, when $\qty(\mathcal{A}_{\mathrm{oct+cq,(2B)}}^{+})_{\omega}/\qty(\mathcal{A}_{\mathrm{oct+cq,(2B)}}^{+})_{3\omega} = 3/9$, which corresponds to the case where the detector views the binary directly face-on ($\iota_{\mathrm{(2B)}} = 0^{\circ}$), then $\min(\beta_1) = 1/6$. However, in this case, $\iota_{\mathrm{(2B)}} = 0^{\circ} = \iota_{\mathrm{(3B)}}$, and the 0.5PN waveform of both the binary and the Lagrange triple vanishes. On the other hand, if $\qty(\mathcal{A}_{\mathrm{oct+cq,(2B)}}^{+})_{\omega}/\qty(\mathcal{A}_{\mathrm{oct+cq,(2B)}}^{+})_{3\omega} = 5/9$, which corresponds to the case where the detector is in the orbital plane of the binary, then $\max(\beta_1) = 1/6$. In this case, $\iota_{\mathrm{(2B)}} = 90^{\circ} = \iota_{\mathrm{(3B)}}$, and the cross modes of the 0.5PN waveforms of both systems also vanish.
            
            We now question the stability of the Lagrange triple satisfying this degeneracy. Referring back to \eqref{eq:3B_stability2}, we see that $\beta_1 = 1/6$ corresponds to an unstable Lagrange triple. In general, even if a binary and a Lagrange triple are degenerate up to the 0.5PN in either of the polarization modes, none of these Lagrange triples are stable under linear perturbations since the range of values of $\beta_1$ as shown in equations \eqref{eq:octDegeneracy_minBeta}--\eqref{eq:octDegeneracy_maxBeta} is well beyond the stability region in equation \eqref{eq:3B_stability2}. This is also shown in the plot of $\iota_{\mathrm{(3B)}}$ in Figure \ref{fig:iotaVsMassRatios}.
            
            In a sense, the amplitude of the 0.5PN waveforms gives a distinguishability criterion for astrophysical binary systems and Lagrange three-body systems. Since astrophysical Lagrange triples must be at least linearly stable, one can look at the 0.5PN waveforms to determine if the observed system is a binary or a Lagrange triple. In particular, the amplitude of the 0.5PN waveform of a Lagrange triple is usually larger than the amplitude of the 0.5PN waveform of a symmetric binary since the former does not simply depend on the mass differences.

        \subsection{On degeneracy and orbits of the Lagrange triple}
            Mathematically, waveform degeneracy exists because there are solutions to the system of equations dictating the equality of the amplitudes and phase. We comment on a deeper reason for waveform degeneracy in terms of the orbits of the Lagrange triple. In the COM frame, the equations of motion of the Lagrange triple can be written as
        	\begin{equation}
        		\ddot{\vb{X}}_{i,\mathrm{(3B)}} = -\frac{M_{i,\mathrm{(3B)}}^{\mathrm{eff}}}{X_{i,\mathrm{(3B)}}^3} \vb{X}_{i,\mathrm{(3B)}}, \qquad i = 1,2,3,
        	\end{equation}
        	where $M_{i,\mathrm{(3B)}}^{\mathrm{eff}}$ is called the effective mass and the orbital radius $X_{i,\mathrm{(3B)}}$ of the $i$th mass are \cite{torigoe_2009, asada_2009}:
        	\begin{align}
        		M_{i,\mathrm{(3B)}}^{\mathrm{eff}} &= \frac{ \qty(m_{j,\mathrm{(3B)}}^2 + m_{k,\mathrm{(3B)}}^2 + m_{j,\mathrm{(3B)}} m_{k,\mathrm{(3B)}})^{3/2} }{M_{\mathrm{(3B)}}^2}, \\
                X_{i,\mathrm{(3B)}} &= b \sqrt{\beta_j^2 + \beta_k^2 + \beta_j\beta_k},
        	\end{align}
            for $i, j, k$ cyclic. If $m_{1,\mathrm{(3B)}} = m_{2,\mathrm{(3B)}}$, then $M_{1,\mathrm{(3B)}}^{\mathrm{eff}} = M_{2,\mathrm{(3B)}}^{\mathrm{eff}}$ and $X_{1,\mathrm{(3B)}} = X_{2,\mathrm{(3B)}}$, meaning that the equal masses move along the same quasicircular orbit. Owing to the triangle configuration, the equal masses maintain a constant angular displacement with each other. Hence, the constraint $m_{1,\mathrm{(3B)}} = m_{2,\mathrm{(3B)}}$ not only removes the time lags between the mass quadrupole and the 0.5PN waveforms but also reduces the number of independent orbits describing the Lagrange triple from three down to two. Since the waveforms depend on the trajectory of the masses, the fact that both the binary and the Lagrange triple with $m_{1,\mathrm{(3B)}} = m_{2,\mathrm{(3B)}}$ being described by two quasicircular orbits then gives rise to the possibility of waveform degeneracy.
    
	\section{Conclusion}\label{sec:conclusion}
		When considering only binary systems and Lagrange three-body systems, the observable time lags between the mass quadrupole waveform and the 0.5PN waveform disentangle the two systems. In particular, if the observable time lags satisfy equations \eqref{eq:3B_timeLags_omega} and \eqref{eq:3B_timeLags_3omega}, then the source must be a Lagrange triple \cite{asada_2009}. Otherwise, it could have come from other systems. In this paper, we showed that when these time lags vanish, the waveform of a binary can still be confused with that of a Lagrange triple.
		
		Specifically, looking at the mass quadrupole alone, we found that there are binary systems and linearly stable Lagrange three-body systems that have the same mass quadrupole waveform in both the plus and cross modes at short timescales. The degeneracy space is parametrized by two parameters, which in this work, we chose to be the total mass $M_{\mathrm{(3B)}}$ and the normalized mass $\beta_1$. The question of which system coalesces first depends on the chirp mass. In particular, the system with the greater chirp mass coalesces first. We also found that there are linearly stable Lagrange triples that can have the same chirp mass as that of a given binary, thus making them have the same mass quadrupole waveform up to the coalescence time. For binaries whose mass difference is small, the waveform match with respect to the noise spectral density of the advanced LIGO design remains above 0.97. This result further highlights the need for including higher PN contributions to modeling binaries and three-body systems. We also showed that there exists waveform degeneracy up to the 0.5PN order, and that this degeneracy space consists of only one point. However, the Lagrange triple that satisfies this degeneracy is unstable.
        
        One immediate extension is to include nonzero eccentricities $e$ to the binary and the Lagrange triple. For an eccentric binary system, there will be additional terms in the amplitude of the plus and cross modes that are proportional to $e$ and $e^2$ \cite{jamet_2025}. We expect similar terms in the waveforms of the elliptic Lagrange triple. Additionally, not only does the semimajor axis of the orbit of each mass change with time, but also the eccentricity. Because of these modifications, the analytical expressions for waveform degeneracy we derived here may not hold for eccentric systems. Of course, only when the GWs have circularized the elliptic orbits enough do our results hold. It will be interesting to analyze waveform degeneracy between eccentric systems since there are values of the normalized masses that make the elliptic Lagrange triple stable even if its circular counterpart is unstable \cite{roberts_2002}. One may also study whether there are waveform degeneracies between elliptic binary systems and other periodic and linearly stable three-body systems in which the masses have close encounters, e.g., figure eight systems.
	
	\begin{acknowledgments}
		The authors would like to thank Jezreel Castillo for feedback on an earlier draft of this paper. CJD is supported by the DOST-SEI through their Undergraduate Scholarship Program.
	\end{acknowledgments}
	
	\section*{Data Availability}
		The data that support the findings of this study were generated by numerical simulations. The source code and parameters used to generate the simulations are publicly available \cite{degeneracy}.
	
	\bibliography{apssamp}

@article{asada_2009,
  title = {Gravitational wave forms for a three-body system in Lagrange's orbit: Parameter determinations and a binary source test},
  author = {Asada, Hideki},
  journal = {Phys. Rev. D},
  volume = {80},
  issue = {6},
  pages = {064021},
  numpages = {10},
  year = {2009},
  month = {Sep},
  publisher = {American Physical Society},
  doi = {10.1103/PhysRevD.80.064021},
  url = {https://link.aps.org/doi/10.1103/PhysRevD.80.064021}
}

@article{torigoe_2009,
  title = {Gravitational Wave Forms for Two- and Three-Body Gravitating Systems},
  author = {Torigoe, Yuji and Hattori, Keisuke and Asada, Hideki},
  journal = {Phys. Rev. Lett.},
  volume = {102},
  issue = {25},
  pages = {251101},
  numpages = {4},
  year = {2009},
  month = {Jun},
  publisher = {American Physical Society},
  doi = {10.1103/PhysRevLett.102.251101},
  url = {https://link.aps.org/doi/10.1103/PhysRevLett.102.251101}
}

@article{liu_2021,
	title = {Simulation of the orbit and spin period evolution of the double pulsars {PSR} {J0737}-3039 from their birth to coalescence induced by gravitational wave radiation},
	volume = {21},
	issn = {1674-4527},
	url = {https://dx.doi.org/10.1088/1674-4527/21/4/104},
	doi = {10.1088/1674-4527/21/4/104},
	number = {4},
	journal = {Res. Astron. Astrophys.},
	author = {Liu, Peng and Yang, Yi-Yan and Zhang, Jian-Wei and Rah, Maria},
	month = may,
	year = {2021},
	pages = {104},
}

@article{blanchet_2014,
	title = {Gravitational {Radiation} from {Post}-{Newtonian} {Sources} and {Inspiralling} {Compact} {Binaries}},
	volume = {17},
	url = {https://doi.org/10.12942/lrr-2014-2},
	doi = {10.12942/lrr-2014-2},
	number = {1},
	journal = {Living Rev. Relativ.},
	author = {Blanchet, Luc},
	month = feb,
	year = {2014},
	pages = {2},
}

@article{musielak_2014,
	title = {The three-body problem},
	volume = {77},
	url = {https://iopscience.iop.org/article/10.1088/0034-4885/77/6/065901},
	doi = {10.1088/0034-4885/77/6/065901},
	number = {6},
	journal = {Rep. Prog. Phys.},
	author = {Musielak, Z E and Quarles, B},
	month = jun,
	year = {2014},
	pages = {065901},
}

@article{essen_2000,
	title = {On the equilateral triangle solution to the three-body problem},
	volume = {21},
	issn = {0143-0807},
	url = {https://dx.doi.org/10.1088/0143-0807/21/6/309},
	doi = {10.1088/0143-0807/21/6/309},
	number = {6},
	journal = {Eur. J. Phys.},
	author = {Essén, Hanno},
	month = nov,
	year = {2000},
	pages = {579},
}

@article{mansilla_2006,
	title = {Stability of {Hamiltonian} {Systems} with {Three} {Degrees} of {Freedom} and the {Three} {Body}-{Problem}},
	volume = {94},
	issn = {1572-9478},
	url = {https://doi.org/10.1007/s10569-005-5360-6},
	doi = {10.1007/s10569-005-5360-6},
	number = {3},
	journal = {Celestial Mech Dyn Astr},
	author = {Mansilla, José Edmundo},
	month = mar,
	year = {2006},
	pages = {249--269},
}

@book{maggiore_vol1,
	title = {Gravitational {Waves}: {Volume} 1: {Theory} and {Experiments}},
	isbn = {978-0-19-857074-5},
	shorttitle = {Gravitational {Waves}},
	url = {https://doi.org/10.1093/acprof:oso/9780198570745.001.0001},
	publisher = {Oxford University Press},
	author = {Maggiore, Michele},
	month = oct,
	year = {2007},
	doi = {10.1093/acprof:oso/9780198570745.001.0001},
}

@article{GW170817_2017,
	title = {{GW170817}: {Observation} of {Gravitational} {Waves} from a {Binary} {Neutron} {Star} {Inspiral}},
	volume = {119},
	shorttitle = {{GW170817}},
	url = {https://link.aps.org/doi/10.1103/PhysRevLett.119.161101},
	doi = {10.1103/PhysRevLett.119.161101},
	number = {16},
	journal = {Phys. Rev. Lett.},
	author = {Abbott, B. P. and others},
	collaboration = {LIGO Scientific and Virgo Collaborations},
	month = oct,
	year = {2017},
	pages = {161101},
}

@book{creighton_2011,
	title = {Gravitational‐{Wave} {Physics} and {Astronomy}: {An} {Introduction} to {Theory}, {Experiment} and {Data} {Analysis}},
	doi = {10.1002/9783527636037},
	publisher = {John Wiley \& Sons, Ltd},
	author = {Creighton, J. D. E. and Anderson, W. G.},
	year = {2011},
}

@book{misner_1973,
	author = {Misner, Charles W. and Thorne, K. S. and Wheeler, J. A.},
	title = {{Gravitation}},
	isbn = {978-0-7167-0344-0, 978-0-691-17779-3},
	publisher = {W. H. Freeman},
	address = {San Francisco},
	year = {1973}
}

@article{roberts_2002,
	title = {Linear {Stability} of the {Elliptic} {Lagrangian} {Triangle} {Solutions} in the {Three}-{Body} {Problem}},
	volume = {182},
	issn = {0022-0396},
	url = {https://www.sciencedirect.com/science/article/pii/S0022039601940896},
	doi = {10.1006/jdeq.2001.4089},
	number = {1},
	journal = {Journal of Differential Equations},
	author = {Roberts, Gareth E.},
	month = jun,
	year = {2002},
	pages = {191--218},
}

@misc{gwastropycbc_2024,
	title = {gwastro/pycbc: v2.3.3 release of {PyCBC}},
	shorttitle = {gwastro/pycbc},
	url = {https://zenodo.org/records/10473621},
	publisher = {Zenodo},
	author = {Nitz, Alex and others},
	month = jan,
	year = {2024},
	doi = {10.5281/zenodo.10473621},
}

@article{abbott_2009,
	title = {{LIGO}: the {Laser} {Interferometer} {Gravitational}-{Wave} {Observatory}},
	volume = {72},
	issn = {0034-4885},
	shorttitle = {{LIGO}},
	url = {https://doi.org/10.1088/0034-4885/72/7/076901},
	doi = {10.1088/0034-4885/72/7/076901},
	number = {7},
	journal = {Rep. Prog. Phys.},
	author = {Abbott, B P and others ({LIGO Scientific and Virgo Collaborations})},
	month = jun,
	year = {2009},
	pages = {076901},
}

@article{ligo_observation_2016,
	title = {Observation of {Gravitational} {Waves} from a {Binary} {Black} {Hole} {Merger}},
	volume = {116},
	url = {https://link.aps.org/doi/10.1103/PhysRevLett.116.061102},
	doi = {10.1103/PhysRevLett.116.061102},
	number = {6},
	journal = {Phys. Rev. Lett.},
	author = {Abbott, B. and others},
	collaboration = {LIGO Scientific and Virgo Collaborations},
	month = feb,
	year = {2016},
	pages = {061102},
}

@article{GW190412_2020,
	title = {{GW190412}: {Observation} of a binary-black-hole coalescence with asymmetric masses},
	volume = {102},
	issn = {2470-0010, 2470-0029},
	shorttitle = {{GW190412}},
	url = {https://link.aps.org/doi/10.1103/PhysRevD.102.043015},
	doi = {10.1103/PhysRevD.102.043015},
	number = {4},
	journal = {Phys. Rev. D},
	author = {Abbott, R. and others},
	collaboration = {LIGO Scientific and Virgo Collaborations},
	month = aug,
	year = {2020},
	pages = {043015},
}

@article{GW190814_2020,
	title = {{GW190814}: {Gravitational} {Waves} from the {Coalescence} of a 23 {Solar} {Mass} {Black} {Hole} with a 2.6 {Solar} {Mass} {Compact} {Object}},
	volume = {896},
	issn = {2041-8205},
	shorttitle = {{GW190814}},
	url = {https://doi.org/10.3847/2041-8213/ab960f},
	doi = {10.3847/2041-8213/ab960f},
	number = {2},
	journal = {ApJL},
	author = {Abbott, R. and others},
	collaboration = {LIGO Scientific and Virgo Collaborations},
	month = jun,
	year = {2020},
	pages = {L44},
}

@article{GW170814_2017,
	title = {{GW170814}: {A} {Three}-{Detector} {Observation} of {Gravitational} {Waves} from a {Binary} {Black} {Hole} {Coalescence}},
	volume = {119},
	shorttitle = {{GW170814}},
	url = {https://link.aps.org/doi/10.1103/PhysRevLett.119.141101},
	doi = {10.1103/PhysRevLett.119.141101},
	number = {14},
	journal = {Phys. Rev. Lett.},
	author = {Abbott, B. P. and others},
	collaboration = {LIGO Scientific and Virgo Collaborations},
	month = oct,
	year = {2017},
	pages = {141101},
}

@article{barandiaran_2024,
	title = {Gravitational waves in the circular restricted three body problem},
	volume = {41},
	issn = {0264-9381},
	url = {https://doi.org/10.1088/1361-6382/ad36a7},
	doi = {10.1088/1361-6382/ad36a7},
	number = {9},
	journal = {Class. and Quantum Grav.},
	author = {Barandiaran, Mikel Martin and Kuroyanagi, Sachiko and Nesseris, Savvas},
	month = apr,
	year = {2024},
	pages = {095002},
}

@article{yamada_2012,
	title = {Triangular solution to the general relativistic three-body problem for general masses},
	volume = {86},
	copyright = {http://link.aps.org/licenses/aps-default-license},
	issn = {1550-7998, 1550-2368},
	url = {https://link.aps.org/doi/10.1103/PhysRevD.86.124029},
	doi = {10.1103/physrevd.86.124029},
	number = {12},
	journal = {Phys. Rev. D},
	publisher = {American Physical Society (APS)},
	author = {Yamada, Kei and Asada, Hideki},
	month = dec,
	year = {2012},
}

@article{ichita_2011,
	title = {Post-{Newtonian} effects on {Lagrange}’s equilateral triangular solution for the three-body problem},
	volume = {83},
	issn = {1550-7998, 1550-2368},
	url = {https://link.aps.org/doi/10.1103/PhysRevD.83.084026},
	doi = {10.1103/PhysRevD.83.084026},
	number = {8},
	journal = {Phys. Rev. D},
	author = {Ichita, Takumi and Yamada, Kei and Asada, Hideki},
	month = apr,
	year = {2011},
	pages = {084026},
}

@article{yamada_2015,
	title = {Post-{Newtonian} effects on the stability of the triangular solution in the three-body problem for general masses},
	volume = {91},
	copyright = {http://link.aps.org/licenses/aps-default-license},
	issn = {1550-7998, 1550-2368},
	url = {https://link.aps.org/doi/10.1103/PhysRevD.91.124016},
	doi = {10.1103/PhysRevD.91.124016},
	number = {12},
	journal = {Phys. Rev. D},
	author = {Yamada, Kei and Tsuchiya, Takuya and Asada, Hideki},
	month = jun,
	year = {2015},
	pages = {124016},
}

@article{yamada_2016,
	title = {Nonchaotic evolution of triangular configuration due to gravitational radiation reaction in the three-body problem},
	volume = {93},
	url = {https://link.aps.org/doi/10.1103/PhysRevD.93.084027},
	doi = {10.1103/PhysRevD.93.084027},
	number = {8},
	journal = {Phys. Rev. D},
	publisher = {American Physical Society},
	author = {Yamada, Kei and Asada, Hideki},
	month = apr,
	year = {2016},
	pages = {084027},
}

@article{sicardy_2010,
	title = {Stability of the triangular {Lagrange} points beyond {Gascheau}’s value},
	volume = {107},
	issn = {1572-9478},
	url = {https://doi.org/10.1007/s10569-010-9259-5},
	doi = {10.1007/s10569-010-9259-5},
	number = {1},
	journal = {Celestial Mechanics and Dynamical Astronomy},
	author = {Sicardy, B.},
	month = jun,
	year = {2010},
	pages = {145--155},
}

@book{poisson_2014,
	address = {Cambridge},
	title = {Gravity: {Newtonian}, {Post}-{Newtonian}, {Relativistic}},
	isbn = {978-1-107-03286-6},
	publisher = {Cambridge University Press},
	author = {Poisson, Eric and Will, Clifford M.},
	year = {2014},
	doi = {10.1017/CBO9781139507486},
}

@article{blanchet_2024,
	title = {Post-{Newtonian} theory for gravitational waves},
	volume = {27},
	url = {https://doi.org/10.1007/s41114-024-00050-z},
	doi = {10.1007/s41114-024-00050-z},
	number = {1},
	journal = {Living Rev. Relativ.},
	author = {Blanchet, Luc},
	month = jul,
	year = {2024},
	pages = {4},
}

@article{pati_2000,
	title = {Post-{Newtonian} gravitational radiation and equations of motion via direct integration of the relaxed {Einstein} equations: {Foundations}},
	volume = {62},
	url = {https://link.aps.org/doi/10.1103/PhysRevD.62.124015},
	doi = {10.1103/PhysRevD.62.124015},
	number = {12},
	journal = {Phys. Rev. D},
	author = {Pati, Michael E. and Will, Clifford M.},
	month = nov,
	year = {2000},
	pages = {124015},
}

@article{dmitrasinovic_2014,
	title = {Gravitational {Waves} from {Periodic} {Three}-{Body} {Systems}},
	volume = {113},
	url = {https://link.aps.org/doi/10.1103/PhysRevLett.113.101102},
	doi = {10.1103/PhysRevLett.113.101102},
	number = {10},
	journal = {Phys. Rev. Lett.},
	author = {Dmitrašinović, V. and Šuvakov, Milovan and Hudomal, Ana},
	month = sep,
	year = {2014},
	pages = {101102},
}

@article{klein_2009,
	title = {Parameter estimation for coalescing massive binary black holes with {LISA} using the full 2-post-{Newtonian} gravitational waveform and spin-orbit precession},
	volume = {80},
	url = {https://link.aps.org/doi/10.1103/PhysRevD.80.064027},
	doi = {10.1103/PhysRevD.80.064027},
	number = {6},
	journal = {Phys. Rev. D},
	author = {Klein, Antoine and Jetzer, Philippe and Sereno, Mauro},
	month = sep,
	year = {2009},
	pages = {064027},
}

@article{owen_1996,
	title = {Search templates for gravitational waves from inspiraling binaries: {Choice} of template spacing},
	volume = {53},
	shorttitle = {Search templates for gravitational waves from inspiraling binaries},
	number = {12},
	journal = {Phys. Rev. D},
	author = {Owen, Benjamin J.},
	month = jun,
	year = {1996},
	pages = {6749--6761},
}

@article{moore_1993,
	title = {Braids in classical dynamics},
	volume = {70},
	doi = {10.1103/PhysRevLett.70.3675},
	number = {24},
	journal = {Phys. Rev. Lett.},
	author = {Moore, Cristopher},
	year = {1993},
	pages = {3675--3679},
}

@article{henon_1976,
	title = {A family of periodic solutions of the planar three-body problem, and their stability},
	volume = {13},
	url = {https://doi.org/10.1007/BF01228647},
	doi = {10.1007/BF01228647},
	number = {3},
	journal = {Celestial mechanics},
	author = {Hénon, M.},
	month = may,
	year = {1976},
	pages = {267--285},
}

@article{moore_2006,
	title = {New {Periodic} {Orbits} for the n-{Body} {Problem}},
	volume = {1},
	issn = {1555-1415},
	url = {https://doi.org/10.1115/1.2338323},
	doi = {10.1115/1.2338323},
	number = {4},
	journal = {J. Comput. Nonlinear Dynam},
	author = {Moore, Cristopher and Nauenberg, Michael},
	month = mar,
	year = {2006},
	pages = {307--311},
}

@article{li_2021,
	title = {Figure-eight orbits in three post-{Newtonian} formulations of triple black holes},
	volume = {104},
	url = {https://link.aps.org/doi/10.1103/PhysRevD.104.044039},
	doi = {10.1103/PhysRevD.104.044039},
	number = {4},
	journal = {Phys. Rev. D},
	publisher = {American Physical Society},
	author = {Li, Dan and Wu, Xin and Liang, Enwei},
	month = aug,
	year = {2021},
	pages = {044039},
}

@article{galaviz_2011,
	title = {Characterization of the gravitational wave emission of three black holes},
	volume = {83},
	url = {https://link.aps.org/doi/10.1103/PhysRevD.83.084013},
	doi = {10.1103/PhysRevD.83.084013},
	number = {8},
	journal = {Phys. Rev. D},
	publisher = {American Physical Society},
	author = {Galaviz, Pablo and Brügmann, Bernd},
	month = apr,
	year = {2011},
	pages = {084013},
}

@article{imai_2007,
	title = {Choreographic {Solution} to the {General}-{Relativistic} {Three}-{Body} {Problem}},
	volume = {98},
	url = {https://link.aps.org/doi/10.1103/PhysRevLett.98.201102},
	doi = {10.1103/PhysRevLett.98.201102},
	number = {20},
	journal = {Phys. Rev. Lett.},
	publisher = {American Physical Society},
	author = {Imai, Tatsunori and Chiba, Takamasa and Asada, Hideki},
	month = may,
	year = {2007},
	pages = {201102},
}

@article{gascheau_1843,
	title={Examen d’une classe d’{\'e}quations diff{\'e}rentielles et applicationa un cas particulier du probleme des trois corps},
	author={Gascheau, M},
	journal={Comptes Rendus},
	volume={16},
	number={7},
	pages={393},
	year={1843}
}

@article{mahapatra_2024,
	title = {Octupolar test of general relativity},
	volume = {109},
	url = {https://link.aps.org/doi/10.1103/PhysRevD.109.024050},
	doi = {10.1103/PhysRevD.109.024050},
	number = {2},
	journal = {Phys. Rev. D},
	publisher = {American Physical Society},
	author = {Mahapatra, Parthapratim},
	month = jan,
	year = {2024},
	pages = {024050},
}

@book{gurfill_2016,
	address = {Berlin, Heidelberg},
	series = {Astrophysics and {Space} {Science} {Library}},
	title = {Celestial {Mechanics} and {Astrodynamics}: {Theory} and {Practice}},
	volume = {436},
	isbn = {978-3-662-50368-3 978-3-662-50370-6},
	shorttitle = {Celestial {Mechanics} and {Astrodynamics}},
	url = {http://link.springer.com/10.1007/978-3-662-50370-6},
	doi = {10.1007/978-3-662-50370-6},
	publisher = {Springer},
	author = {Gurfil, Pini and Seidelmann, P. Kenneth},
	year = {2016},
}

@article{nakamura_2023,
	title = {Collinear and triangular solutions to the coplanar and circular three-body problem in the parametrized post-{Newtonian} formalism},
	volume = {107},
	issn = {2470-0010, 2470-0029},
	url = {https://link.aps.org/doi/10.1103/PhysRevD.107.044005},
	doi = {10.1103/PhysRevD.107.044005},
	number = {4},
	journal = {Phys. Rev. D},
	author = {Nakamura, Yuya and Asada, Hideki},
	month = feb,
	year = {2023},
	pages = {044005},
}

@article{nakamura_2024,
	title = {Triangular solution to the planar elliptic three-body problem in the parametrized post-{Newtonian} formalism},
	volume = {109},
	issn = {2470-0010, 2470-0029},
	url = {https://link.aps.org/doi/10.1103/PhysRevD.109.064067},
	doi = {10.1103/PhysRevD.109.064067},
	number = {6},
	journal = {Phys. Rev. D},
	author = {Nakamura, Yuya and Asada, Hideki},
	month = mar,
	year = {2024},
	pages = {064067},
}

@misc{lalsuite,
       author         = "{LIGO Scientific Collaboration} and {Virgo Collaboration} and {KAGRA Collaboration}",
       title          = "{LVK} {A}lgorithm {L}ibrary - {LALS}uite",
       howpublished   = "Free software (GPL)",
       doi            = "10.7935/GT1W-FZ16",
       year           = "2018"
 }

@article{schnittman_2010,
	title = {{THE} {LAGRANGE} {EQUILIBRIUM} {POINTS} {L4} {AND} {L5} {IN} {BLACK} {HOLE} {BINARY} {SYSTEM}},
	volume = {724},
	issn = {0004-637X},
	url = {https://doi.org/10.1088/0004-637X/724/1/39},
	doi = {10.1088/0004-637X/724/1/39},
	number = {1},
	journal = {ApJ},
	publisher = {The American Astronomical Society},
	author = {Schnittman, Jeremy D.},
	month = oct,
	year = {2010},
	pages = {39},
}

@article{seto_2010,
	title = {Relativistic astrophysics with resonant multiple inspirals},
	volume = {81},
	url = {https://link.aps.org/doi/10.1103/PhysRevD.81.103004},
	doi = {10.1103/PhysRevD.81.103004},
	number = {10},
	journal = {Phys. Rev. D},
	publisher = {American Physical Society},
	author = {Seto, Naoki and Muto, Takayuki},
	month = may,
	year = {2010},
	pages = {103004},
}

@article{jamet_2025,
	title = {Emission and detection of ultrahigh frequency gravitational waves from highly eccentric orbits of compact binary systems},
	volume = {111},
	url = {https://link.aps.org/doi/10.1103/PhysRevD.111.103032},
	doi = {10.1103/PhysRevD.111.103032},
	number = {10},
	journal = {Phys. Rev. D},
	publisher = {American Physical Society},
	author = {Jamet, Pierre and Barrau, Aurélien and Martineau, Killian},
	year = {2025},
	pages = {103032},
}

@article{morbidelli_2005,
	title = {Chaotic capture of {Jupiter}'s {Trojan} asteroids in the early {Solar} {System}},
	volume = {435},
	copyright = {2005 Macmillan Magazines Ltd.},
	url = {https://www.nature.com/articles/nature03540},
	doi = {10.1038/nature03540},
	number = {7041},
	journal = {Nature},
	publisher = {Nature Publishing Group},
	author = {Morbidelli, A. and Levison, H. F. and Tsiganis, K. and Gomes, R.},
	month = may,
	year = {2005},
	pages = {462--465},
}

@article{stegmann_2022,
	title = {Binary black hole mergers from merged stars in the {Galactic} field},
	volume = {106},
	url = {https://link.aps.org/doi/10.1103/PhysRevD.106.023014},
	doi = {10.1103/PhysRevD.106.023014},
	number = {2},
	journal = {Phys. Rev. D},
	publisher = {American Physical Society},
	author = {Stegmann, Jakob and Antonini, Fabio and Schneider, Fabian R. N. and Tiwari, Vaibhav and Chattopadhyay, Debatri},
	month = jul,
	year = {2022},
	pages = {023014},
}

@article{romero-shaw_2021,
	title = {Signs of {Eccentricity} in {Two} {Gravitational}-wave {Signals} {May} {Indicate} a {Subpopulation} of {Dynamically} {Assembled} {Binary} {Black} {Holes}},
	volume = {921},
	url = {https://doi.org/10.3847/2041-8213/ac3138},
	doi = {10.3847/2041-8213/ac3138},
	number = {2},
	journal = {ApJL},
	publisher = {The American Astronomical Society},
	author = {Romero-Shaw, Isobel and Lasky, Paul D. and Thrane, Eric},
	month = nov,
	year = {2021},
	pages = {L31},
}

@article{zwart_2026,
	title = {The formation of periodic three-body orbits for {Newtonian} systems},
	volume = {707},
	url = {https://www.aanda.org/articles/aa/abs/2026/03/aa58230-25/aa58230-25.html},
	doi = {10.1051/0004-6361/202558230},
	journal = {A\&A},
	publisher = {EDP Sciences},
	author = {Zwart, Simon Portegies and Doelman, Arjen and Sein, Jelmer},
	month = mar,
	year = {2026},
	pages = {A215},
}

@misc{degeneracy,
	title = {{BinaryLagrangeDegeneracy}},
	url = {https://doi.org/10.5281/zenodo.21474047},
	publisher = {Zenodo},
	author = {Doctolero, Carlos Jaimel},
	month = jul,
	year = {2026},
	doi = {10.5281/zenodo.21474047},
}
\end{document}